\documentclass[12pt, onecolumn]{article}
\usepackage{amssymb,amsmath}
\usepackage{psfrag}
\newcommand{\newc}{\newcommand}

\def\beq{\begin{equation}}
\def\eeq{\end{equation}}
\def\bea{\begin{eqnarray}}
\def\eea{\end{eqnarray}}
\def\bitem{\begin{itemize}}
\def\eitem{\end{itemize}}
\newc{\ie}{{\it i.e. }}          \newc{\etal}{{\it et al. }}
\newc{\eg}{{\it e.g. }}          \newc{\etc}{{\it etc. }}
\newc{\cf}{{\it c.f. }}
\def\bar#1{\overline{#1}}

\def\inv{^{\raise.15ex\hbox{${\scriptscriptstyle -}$}\kern-.05em 1}}
\def\lbar{{\lower.35ex\hbox{$\mathchar'26$}\mkern-10mu\lambda}} 

\def\to{\rightarrow}

\let\p=\partial
\let\<=\langle
\let\>=\rangle

\let\+=\uparrow
\let\-=\downarrow

\newc{\gsim}{\lower.7ex\hbox{$\;\stackrel{\textstyle>}{\sim}\;$}}
\newc{\lsim}{\lower.7ex\hbox{$\;\stackrel{\textstyle<}{\sim}\;$}}
\newc{\gev}{\,{\rm GeV}}
\newc{\mev}{\,{\rm MeV}}
\newc{\ev}{\,{\rm eV}}
\newc{\kev}{\,{\rm keV}}
\newc{\tev}{\,{\rm TeV}}

\newc{\mz}{M_Z}
\newc{\mpl}{M_{\rm pl}}
\newc{\mw}{m_{\rm weak}}
\newc{\mh}{m_{\rm h}}
\newc{\gw}{\hat{g}_2}
\newc{\gy}{\hat{g}_Y}
\newc{\g}{\hat{g}}
\newc{\gst}{\hat{g}_3}
\newc{\lambdah}{\hat{\lambda}}
\newc{\muh}{\hat{\mu}}
\newc{\vh}{\hat{v}}
\newc{\Bkk}{B^{(1)}}
\newc{\Wkk}{W^{(1)}}
\newc{\Wpmkk}{W_{\pm}^{(1)}}
\newc{\Zkk}{W^{3(1)}}
\newc{\gakk}{\gamma^{(1)}}
\newc{\Hkk}{H^{(1)}}
\newc{\hkk}{h^{(1)}}
\newc{\akk}{a_0^{(1)}}
\newc{\apmkk}{a_{\pm}^{(1)}}
\newc{\Bz}{B^{(0)}}
\newc{\Wz}{W^{(0)}}
\newc{\Wpmz}{W_{\pm}^{(0)}}
\newc{\Zz}{B^{(0)}}
\newc{\gaz}{\gamma^{(0)}}
\newc{\Hz}{H^{(0)}}
\newc{\hz}{h^{(0)}}
\newc{\mwh}{\hat{m}_W}
\newc{\mbh}{\hat{m}_B}
\newc{\mzh}{\hat{m}_Z}
\begin{document}
\thispagestyle{empty}
\vspace*{.5cm}
\noindent
\hspace*{\fill}{\large MCTP-08-64}\\
\vspace*{2.0cm}

\begin{center}
{\Large\bf Non-minimal universal extra dimensions}
\\[2.5cm]
{\large Thomas Flacke$^a$, A. Menon$^a$, and Daniel J. Phalen$^a$
}\\[.5cm]
{\it $^a$ Michigan Center for Theoretical Physics (MCTP)\\Randall Laboratory,
Physics Department, University of Michigan\\Ann Arbor, MI 48109 }
\\[.2cm]
(Dated: \today)
\\[1.1cm]

{\bf Abstract}\end{center}
\noindent
In this paper we investigate the phenomenological implications of boundary
localized terms (BLTs) in the model of Universal Extra Dimensions (UED).
In particular, we study the electroweak Kaluza-Klein mass spectrum resulting
from BLTs and their effect on electroweak symmetry breaking
via the five dimensional Higgs mechanism. We find that the addition of BLTs to massive five dimensional fields induces a non-trivial extra dimensional profile for the zero and non-zero Kaluza-Klein (KK) modes. Hence BLTs
generically lead to a modification of Standard Model parameters and are
therefore experimentally constrained, even at tree level. We study Standard Model constraints on three representative non-minimal UED models in detail and find that the constraints on BLTs are weak. On the contrary, non-zero BLTs have a major impact on the spectrum and couplings of non-zero KK modes. For
example, there are regions of parameter space where the Lightest Kaluza-Klein
particle (LKP) is either the Kaluza-Klein Higgs boson or the first KK mode of
the $W^{3}$.

\newpage

\setcounter{page}{1}

\section{Introduction}

In models with an Universal Extra Dimension (UED) \cite{Appelquist:2000nn}, all
Standard Model particles are promoted to 5 dimensional fields propagating in a
flat extra dimension.\footnote{For pre-dating ideas closely related to UED models see Refs.~\cite{preUED}.} The extra dimension is chosen to be the orbifold
$S^1/\mathbb{Z}_2$, and with the appropriate boundary conditions one can obtain chiral fermions and avoid the massless scalar
modes associated with the zero modes of the extra gauge field components.
Due to the $\mathbb{Z}_2$ symmetry, the
interactions between the Kaluza-Klein (KK) modes respect a $\mathbb{Z}_2$
parity called KK parity. KK parity implies that Kaluza Klein particles can only be
produced pairwise, leading to a lower bound on the KK mass scale ($M_{kk}$) of about 500~GeV,
which can be probed by the Large Hadron Collider. Furthermore, KK parity
guarantees the stability of the lightest Kaluza-Klein particle (LKP), thus
providing UED with a dark matter candidate.

At tree-level, UED is a simple extension of the Standard Model with only
two experimentally undetermined parameters: the compactification radius $R$
and the Higgs mass $\mh$. However, since UED is a 5 dimensional theory, it is
non-renormalizable and therefore should be considered to be an effective field
theory. The cutoff for UED can be estimated by naive dimensional analysis
to be $\Lambda\sim 50 M_{kk}$~\cite{NDA}, where the Kaluza-Klein mass scale
$M_{kk}\equiv 1/R$ is set by the compactification radius $R$.\footnote{Studies of unitarity bounds on heavy gluon scattering also impose bounds on the number of KK states included in the effective 4D theory, typically implying $\Lambda<\mathcal{O}(10) M_{kk}$ \cite{SekharChivukula:2001hz}.}
Treating UED as an effective field theory implies that all operators
that are allowed by the Standard Model gauge symmetries and 4 dimensional
Lorentz invariance should be included in the theory. Such operators
can be in the bulk or localized at the orbifold fixed points so there
are many more undetermined tree level parameters than $(R,m_h)$.

To our knowledge, all phenomenological UED studies of bounds from
colliders~\cite{UEDcolliderbounds}, electroweak
precision~\cite{UEDprecisionbounds}, flavor changing neutral
currents~\cite{Buras:2002ej,UEDflavorbounds} and other precision
measurements~\cite{UEDprecisionOther} have focused on either the tree level
couplings and mass spectrum of UED without boundary localized terms (which we
will refer to as ``standard'' UED) or on the one loop modified mass spectrum
and couplings of Minimal UED (MUED)~\cite{Cheng:2002iz}. In the MUED scenario,
all boundary localized terms are assumed to be zero at the cutoff scale
$\Lambda$ and are induced
at low scales due to renormalization group evolution. Studies of UED dark
matter~\cite{elastic,Servant:2002aq,UEDDM,Arrenberg:2008wy} often make less
strong assumptions about the details of the KK mass spectrum, but assume that
the couplings of the first KK level excitations are identical to those of the
Standard model. A further important task is to distinguish UED from other
Standard Model extensions \cite{UEDvsSUSY}. If beyond the Standard Model signals
are found at the LHC, they should be studied in the full UED parameter
space.\footnote{For a detailed review on the current status of UED, see
Ref.~\cite{Hooper:2007qk}.}

In this article, we study the phenomenological impact of including boundary
localized terms (BLTs) for the electroweak sector of UED. The electroweak
sector is of
interest because it hosts many of the phenomenologically viable dark matter
candidates: the first KK mode of the $B$ gauge boson $B^{(1)}$,
the first KK mode of the neutral component of the $W$ gauge boson $W^{3(1)}$,
the first KK mode of the Higgs boson $h^{(1)}$ and the electrically neutral
pseudoscalar Higgs boson $a^{0(1)}$.\footnote{The KK neutrino $\nu^{(1)}$ is
experimentally disfavored, if standard UED couplings are assumed. Direct
detection limits on
a KK neutrino scattering off a nucleon through a t-channel Z boson puts a
bound of $M_\nu^{(1)}\gtrsim 50 \tev$ \cite{elastic}, while requiring that
the KK neutrino does not over close the universe
requires $M_\nu^{(1)}\lesssim 3 \tev$ \cite{Servant:2002aq}.} The
BLTs we consider are boundary localized kinetic terms for the $B$ and $W$ gauge
bosons and boundary localized kinetic, mass, and quartic terms for the Higgs
boson. Of the possible additional effective operators that are compatible with
4 dimensional Lorentz invariance and the Standard Model gauge symmetries, these
BLTs have the lowest mass dimension.
We find that the zero mode wavefunctions of the massive 5 dimensional fields
with BLTs are not generally flat, which in UED generically leads to modified
zero mode couplings. As the zero modes of the 5
dimensional fields are to be identified with the Standard Model, these
modifications translate into
constraints on the size of these boundary localized operators. We find these constraints are weak, leaving regions of allowed parameter space where the LKP
is the KK Higgs $h^{(1)}$ and other, extended regions in which the LKP is $W^{3(1)}$-like. We would
like to emphasize
that our analysis is complementary to that of MUED in Ref.~\cite{Cheng:2002iz}
because we assume that the effects of BLTs dominate those
induced by loop effects and our analysis is purely at tree level.

The article is organized as follows: in Section~\ref{secUEDreview} we provide
a brief review of the ``standard'' UED model. We concentrate on the structure
of the KK decomposition and the basic relations between the 4 dimensional and 5
dimensional masses and couplings in order to compare them to the modified
relations established in the latter part of the paper. To include
BLTs in UED we need to KK expand the extra dimensional
fields in the presence of BLTs and extend the standard UED gauge fixing
procedure of Ref.~\cite{Muck:2001yv} to identify the physical Higgs and
Goldstone modes at each KK level.

In Section~\ref{secmassiveBLKT} we consider the toy model of
a massive 5 dimensional scalar field on $S_1/\mathbb{Z}_2$ with boundary
localized kinetic terms (BLKTs) and boundary localized mass terms (BMTs) to demonstrate the effect of BLTs on the wavefunctions of an extra
dimensional field.
We find that the mass spectrum and wavefunctions are significantly modified
by BLTs. In particular, the wavefunction of the zero mode becomes
non-flat. The transcendental equations that determine the scalar
KK mode masses in Ref.~\cite{Dvali:2001gm} are modified due to the
additional 5 dimensional bulk mass parameter. These KK mode mass relations,
wavefunctions and normalizations remain the same for gauge fields and hence
can be translated directly into those for the electroweak sector. A
generalization of the UED gauge fixing procedure necessary for the
identification of the boundary conditions of the physical Higgses $a^\pm$ and
$a^0$ is worked out in the Appendices~\ref{appgaugefix} and~\ref{appgaugefixbd}.
In particular we determine the Goldstone's, pseudoscalar's, and charged Higgs
bosons' equations of motion and their boundary conditions in unitary
gauge. From their boundary conditions we are able to determine the mass
spectrum of the pseudoscalars and their couplings.

In Section~\ref{secUEDtheory} we use these results to present the KK decomposition
of the complete electroweak sector in UED and the modifications of couplings of
the KK modes as well as the zero modes. Using these mass relations and
couplings we discuss the phenomenological consequences of the BLTs in
Section~\ref{secPheno}. We study three sample scenarios in detail in order to
illustrate the constraints and novel phenomenology of non-minimal UED. In
scenario I, we assume uniform electroweak BLKTs and vanishing Higgs boundary
mass and quartic terms, while in scenario II we allow the Higgs BLKT to differ from the uniform gauge BLKT,
and in scenario III we allow the $U(1)_Y$ and $SU(2)$ BLKTs to differ. For each of these
scenarios we match the tree-level zero mode masses and spectra to that of the
Standard Model to constrain the size of the BLTs. For scenario I,
the LKP is the KK photon, while for scenario II there are regions of
allowed parameter space with a Higgs LKP. For scenario III, the LKP is the $W^3$ in most of the parameter space.
Finally in Section~\ref{secConclusion} we conclude.

\section{UED mass spectrum and couplings: a mini review}\label{secUEDreview}

In this section we briefly review the theoretical setup of universal extra
dimensions (UED) and discuss its mass spectrum and couplings in the absence of
large BLTs.\footnote{See Refs.~\cite{Appelquist:2000nn,Hooper:2007qk}.}

\subsection{UED at tree level}\label{secUEDtree}

The UED bulk action on $S^1/\mathbb{Z}_2$ is
\beq
S_{UED,bulk}=S_g+S_H+S_f
\label{UEDbulkaction}
\eeq
with
\bea
S_g &=& \int d^5x \left( -\frac{1}{4\gst^2}G^A_{MN}G^{AMN}-\frac{1}{4\gw^2}
W^I_{MN}W^{IMN}-\frac{1}{4\gy^2}B_{MN}B^{MN} \right)\\
S_H &=& \int d^5x \left( (D_MH)^{\dagger}(D^MH)+\muh^2H^{\dagger}H-\lambdah(
H^{\dagger}H)^2 \right) \\
S_f &=& \int d^5x \left(i \bar{f} \gamma^M D_M f+\left(\lambdah_{E}
\bar{L}E H+\lambdah_{U}\bar{Q}U\tilde{H}+
\lambdah_{D}\bar{Q}D H+\mbox{h.c.}\right)\right)
\eea
where $x_5\equiv y \in [0,\pi R]$, $G_{MN}$, $W_{MN}$, $B_{MN}$ are the 5
dimensional $SU(3)_C\times SU(2)_W\times U(1)_Y$ gauge field strengths,
$f = (Q,U,D,L,E)$ denote the 5 dimensional fermion fields, $D_M$ is the
corresponding 5 dimensional covariant derivative and the hatted quantities
denote the 5 dimensional couplings.

In order to obtain the Standard Model spectrum at the zero mode level,
Neumann boundary conditions are imposed on the $H, G^A_\mu,W^I_\mu,B_\mu,
Q_L,L_L,U_R,D_R,E_R$ fields while Dirichlet boundary conditions are imposed
on the $G^A_5,W^I_5,B_5,Q_R,L_R,U_L,D_L,E_L$ fields.
If no boundary terms are present, all bulk fields can be decomposed in terms of
 the \emph{same} KK mode basis $\{f^{(n)}\}\propto\{\sin(ny/\pi R),\cos(ny/\pi
R)\}$.\footnote{For details on
the fermion decomposition see Ref.~\cite{Buras:2002ej}.} Integration over the
extra dimension yields the effective 4 dimensional action in
terms of the Standard Model and its KK partners. Matching the zero mode masses
and couplings to the Standard Model fixes all 5 dimensional parameters in
terms of the Standard Model observables multiplied by the appropriate factors of $\pi R$..
The only free parameters at tree level are the compactification radius $R$
and the Higgs mass $m_h$.

At the non-zero KK levels, the spectrum contains a partner for every Standard Model
particle with a mass $m_{\Phi^{(n)}}=\sqrt{(n/R)^2+m^2_{\Phi^{(0)}}}$, where
$m^2_{\Phi^{(0)}}$ is the corresponding zero mode mass. In addition to these
fields, at each non-zero KK level a charged Higgs $a^{\pm}$ and a pseudoscalar
Higgs $a^{0}$ are also present. The extra Higgs bosons are present because at
the n$^{\rm th}$ KK level there are eight scalar degrees of freedom due to the
$W_5^{I(n)}$, $B_5^{(n)}$ and the Higgs boson $H^{(n)}$. A linear combination
of four of these scalars form the longitudinal components of the $B_\mu^{(n)}$
and $W_\mu^{a(n)}$ gauge bosons. The remaining four degrees of freedom form the
KK Higgs boson
$h^{(n)}$, the charged Higgs $a^{\pm(n)}$ and the pseudoscalar Higgs $a^{0(n)}
$ with masses $m_{h^{(n)}}=\sqrt{(n/R)^2+m^2_{h^{(0)}}}$, $m_{a^{\pm (n)}}=
m_{W^{\pm(n)}}$, and $m_{a^{0 (n)}}=m_{Z^{(n)}}$. Hence the particle spectrum
at any non-zero KK mode is almost degenerate. A detailed discussion of the
electroweak sector including the identification of the Goldstone and the
physical Higgs mode and their KK decomposition can be found in
Ref.~\cite{Muck:2001yv}.
It is important to note that in the electroweak sector, mixing occurs between
$B^{(n)}_\mu$ and $W^{(n)}_\mu$ as well as in the Goldstone - Higgs sector.
In the absence of BLTs, the KK bases of all fields are identical and therefore
orthogonality guarantees no mixing between different KK-levels.
All non-zero tree-level couplings of the heavier KK modes are the same as those
of the zero mode as long as the vertex satisfies KK number conservation.  KK number is violated at loop level but $\mathbb{Z}_2$ symmetry
guarantees the stability of the lightest KK particle.

\subsection{UED as an effective field theory}\label{secUEDasEFT}

As a five dimensional quantum field theory, UED is non-renormalizable and
should be considered to be an effective field theory. Naive dimensional
analysis suggests that a perturbative description of UED is valid up to the
energy scale $\Lambda\sim 50/R$ \cite{NDA}. Studies of unitarity bounds on heavy gluon scattering also impose bounds on the number of KK states included in the effective 4D theory, typically implying $\Lambda<\mathcal{O}(10) M_{KK}$ \cite{SekharChivukula:2001hz}. Hence for a
phenomenologically interesting compactification radius of $R^{-1} \sim 1$~TeV,
the cutoff of the theory is relatively low.
Without a better understanding of the high scale completion of UED we cannot
{\it a priori} neglect allowed operators, but we should instead try to use
experimental data to put constraints on the size of these operators.
The set of operators that agree with 4 dimensional Lorentz invariance,
$\mathbb{Z}_2$ parity, and the gauge symmetries of the Standard Model include
higher dimensional operators in the bulk and boundary localized operators at
the fixed points of the $S^1/\mathbb{Z}_2$ compactified extra dimension.

Boundary localized operators contain the lowest dimensional operators beyond the ``standard'' UED operators and therefore should be included in any realistic
phenomenological treatment of UED. Furthermore, even if the BLTs
are set to zero at tree-level, they will be generated at the one-loop
level~\cite{Cheng:2002iz,Georgi:2000ks}. In particular, \emph{every} bulk term
in the action in Eq.~(\ref{UEDbulkaction}) can be accompanied by a
corresponding boundary localized operator
at each orbifold fixed point, and the size of these boundary localized operators
needs to be equal due to $\mathbb{Z}_2$ symmetry. The most studied
effective field theoretic description of UED is ``Minimal'' UED
~\cite{Cheng:2002iz}. In MUED there are three undetermined parameters:
the compactification scale $R^{-1}$, the Higgs mass $m_h$ and the cutoff scale
$\Lambda$. All the BLTs are assumed to vanish at the cutoff scale $\Lambda$
but are induced by RG running from $\Lambda$ down to the electroweak scale.
The RG evolution of the bulk and boundary terms induces significant changes
in the particle spectrum and lifts some of the degeneracies of tree-level
UED. One of the most striking features of the one-loop corrected mass
spectrum is that the Weinberg angle differs for different KK levels. The
LKP, being the lighter eigenstate of the $B^{(1)}-W^{3(1)}$ system, turns out
to be almost purely $B^{(1)}$.

In this paper we assume large BLTs at the electroweak
scale and study the impact of including these terms on tree-level Standard
Model observables. This setup is non-minimal because we relax
the assumption of vanishing BLTs at the cutoff scale.

\section{Boundary localized terms for massive bulk fields}\label{secmassiveBLKT}

When compactifying a higher dimensional field on a flat manifold, the KK modes can be understood as discrete eigenstates of momentum in the extra
dimension. KK number conservation in interactions is a remnant of five
dimensional momentum conservation. BLTs violate 5
dimensional translational invariance and therefore induce mixing
between the modes of well defined KK number. If the BLTs are sufficiently suppressed they can be treated as perturbations and dealt with as mass insertions. Since we want to allow for large BLTs, the mass insertions due to BLTs can no longer be treated perturbatively.

The problem of boundary localized kinetic terms (BLKTs) in 5 dimensional
theories has been addressed in Refs.~\cite{Dvali:2001gm,Carena:2002me,delAguila:2003bh}.
In Ref.~\cite{Dvali:2001gm}, it was shown that a 5 dimensional massless scalar
field $\Phi$ with a localized brane kinetic term at $y=0$ of the form
\beq
S_{BLKT}= \frac{r_\Phi}{2} \int d^5x  \; \p_\mu \Phi \p^\mu \Phi \delta(y),
\eeq
can be decomposed into KK modes by demanding that its wavefunctions $f_n$
satisfy the modified orthogonality relations
\bea
\int  dy\left[1+r_\Phi \delta(y)\right] f_n(y)f_m(y)&=&\delta_{nm}
\nonumber\\
\int  dy (\p_5 f_n(y))(\p_5 f_m(y))  &=&  m_{\Phi^{(n)}}^2
\delta_{nm},
\label{orthogonalrelns}
\eea
where $m_{\Phi^{(n)}}$ is the $n^{th}$ KK mode mass.
Ref.~\cite{Carena:2002me} showed that the same prescription works for massless
gauge fields. The inclusion of a second brane at $\pi R$ due to the
$\mathbb{Z}_2$ symmetry just changes $\delta(y)$ to $[\delta(y)+\delta(y-\pi R)
]$ in Eq.~(\ref{orthogonalrelns}). Including BLKTs reduces
the non-zero KK mode masses to values below those of ``standard'' UED. The zero
mode is still massless and zero mode wavefunctions are still flat.

Using these results we can now consider the case of a massive 5 dimensional
scalar which has both boundary localized mass terms and boundary localized
kinetic terms. The bulk action on $S^1/\mathbb{Z}_2$ is
\beq
S=\frac{1}{2} \int d^5 x  \left(\p^M \Phi \p_M \Phi -m^2 \Phi^2 \right),
\eeq
which leads to the equation of motion
\beq
(\Box-\p^2_5+m^2)\Phi=0. \label{PHIEOM}
\eeq
Using separation of variables we decompose the 5 dimensional scalar into the
form
\bea
\Phi(x,y) &=&\sum_i \Phi^{(i)}(x) f_i(y),
\eea
so that the equations of motion are
\bea
\Box \Phi(x) &=& -m_i^2 \Phi^{(i)}(x) \label{Phisep} \\
f_i^{''}(y) &=& -(m_i^2 - m^2) f_i(y) \equiv -M_i^2 f_i(y). \label{fiEOM}
\eea

Due to $S^1/\mathbb{Z}_2$ symmetry, the solutions to Eq.~(\ref{fiEOM}) are
odd or even under $\mathbb{Z}_2$ and are given by
\bea
f_\alpha=N_\alpha \left\{\begin{array}{cc}
\frac{\cosh \left(M_\alpha \left(y-\frac{\pi R}{2}\right)\right)}{\cosh \left(
\frac{M_\alpha \pi R}{2}\right)} & \alpha \mbox{ even} \\
-\frac{\sinh \left(M_\alpha \left(y-\frac{\pi R}{2}\right)\right)}{\sinh \left(
\frac{M_\alpha \pi R}{2}\right)} & \alpha \mbox{ odd} \\
\end{array} \right.
\label{zeroQ}
\eea
\bea
f_n=N_n \left\{\begin{array}{cc}
\frac{\cos \left(M_n \left(y-\frac{\pi R}{2}\right)\right)}{\cos \left(
\frac{M_n \pi R}{2}\right)} & n \mbox{ even} \\
-\frac{\sin \left(M_n \left(y-\frac{\pi R}{2}\right)\right)}{\sin \left(
\frac{M_n \pi R}{2}\right)} & n \mbox{ odd} \\
\end{array} \right.,
\label{phibasis}
\eea
where the physical masses are
\bea
m_\alpha^2&=&-M_\alpha^2+m^2\nonumber\\
m_n^2&=&M_n^2+m^2,\label{Mndef}
\eea
and  we use lower case Greek indices for the hyperbolic solutions and lower
case Latin indices for the trigonometric solutions.\footnote{Depending on the
choice of boundary terms, the hyperbolic equation has zero, one or two
solutions. For the case of a hyperbolic solution the physical mass
$m_{\alpha}^2$ always remains positive.}

So far, our discussion has been independent of the BLTs,
which enter in two ways. First, BLTs modify the normalization conditions that
determine the coefficients $N_\alpha$ and $N_n$. Second, they modify the variation of the action on the boundary.  In order to find a consistent solution, the bulk \emph{and} the boundary variations of
the action must vanish. The boundary variation has contributions from the bulk via partial
integrations in $y$ and directly from the variation of BLTs. Requiring that the boundary variation vanish leads to
boundary conditions on $\Phi$, which result in a quantization condition
on $M_i$.

Adding the following BLTs
\beq
S_{bd}=\frac{1}{2} \int d^5x \left(r_\Phi \p^\mu \Phi \p_\mu \Phi-m_b^2 \Phi^2
\right)
[\delta(y)+\delta(y-\pi R)],
\eeq
 we find the modified boundary conditions
\bea
0&=&[\p_5-(r_\Phi \Box+m_b^2)]\Phi|_{y=0}\\
0&=&[\p_5+(r_\Phi \Box+m_b^2)]\Phi|_{y=\pi R},
\label{phibound}
\eea
where $r_{\Phi}$ is the brane kinetic parameter and $m_b$ is the
brane mass term.
If we use the wavefunctions in Eq.~(\ref{zeroQ}) and Eq.~(\ref{phibasis})
we find the ``hyperbolic'' quantization conditions
\bea
b_\alpha\equiv\frac{(r_{\Phi} m_\alpha^2-m_b^2)}{M_\alpha} = \left\{ \begin{array}{cc}
\tanh (\frac{M_\alpha \pi R}{2}) & \mbox{ even} \\
\coth (\frac{M_\alpha \pi R}{2}) & \mbox{ odd}
\end{array} \right. \label{zeroQcond}
\eea
and the ``trigonometric'' quantization conditions
\bea
b_n\equiv\frac{(r_{\Phi} m_n^2-m_b^2)}{M_n} = \left\{ \begin{array}{cc}
-\tan (\frac{M_n \pi R}{2}) & n \mbox{ even} \\
\cot (\frac{M_n \pi R}{2}) & n \mbox{ odd}.
\end{array} \right. \label{nQcond}
\eea
The wavefunctions $\{f_n\}$, $\{f_\alpha\}$ in Eq.~(\ref{zeroQ}) and
Eq.~(\ref{phibasis}) are pairwise orthonormal with respect to the modified
scalar product
\beq
\int_{0}^{\pi R} dy \left[1+ r_{\Phi} \left[\delta(y) + \delta(y-\pi R) \right]
\right] f_i f_j = \delta_{ij},\label{normcond}
\eeq
resulting in the normalizations
\bea
N_\alpha^{-2} = \left\{\begin{array}{cc}
\mbox{sech}^2 \left(\frac{M_\alpha \pi R}{2} \right) \left[\frac{\sinh(M_\alpha
\pi R)}{2M_\alpha} + \frac{\pi R}{2} \right] + 2 r_{\Phi} & \mbox{even} \\
\mbox{cosech}^2 \left(\frac{M_\alpha \pi R}{2} \right) \left[\frac{\sinh(M_\alpha \pi
R)}{2M_\alpha} - \frac{\pi R}{2} \right] + 2 r_{\Phi} & \mbox{odd}
\end{array} \right.
\label{normcond0}
\eea
\bea
N_n^{-2} = \left\{\begin{array}{cc}
\sec^2 \left(\frac{M_n \pi R}{2} \right) \left[\frac{\pi R}{2} +
\frac{\sin(M_n \pi R)}{2M_n}   \right] + 2 r_{\phi} & \mbox{even} \\
\mbox{cosec}^2 \left(\frac{M_n \pi R}{2} \right) \left[\frac{\pi R}{2} -
\frac{\sin(M_n \pi R)}{2M_n} \right] + 2 r_{\phi} & \mbox{odd}
\end{array} \right. \label{normcondn}.
\eea

Eq.~(\ref{zeroQcond}) and Eq.~(\ref{nQcond}) show the dependence
of the KK masses on the brane kinetic parameter $r_\Phi$ and the brane mass
term $m_b$. If $r_\Phi m^2<m_b^2$, there is no hyperbolic solution, while for
$r_\Phi m^2>m_b^2$ one or two hyperbolic solutions are possible. In addition, a
flat zero mode wavefunction is only possible when
\beq
r_\Phi m^2=m_b^2. \label{flatnesscondPhi}
\eeq
The physical
masses for these solutions interpolate smoothly for different values of
$r_{\Phi}$, $m_b$, $m$ and $R^{-1}$.

\begin{figure}
\begin{center}
\resizebox{7.5cm}{!}{\includegraphics{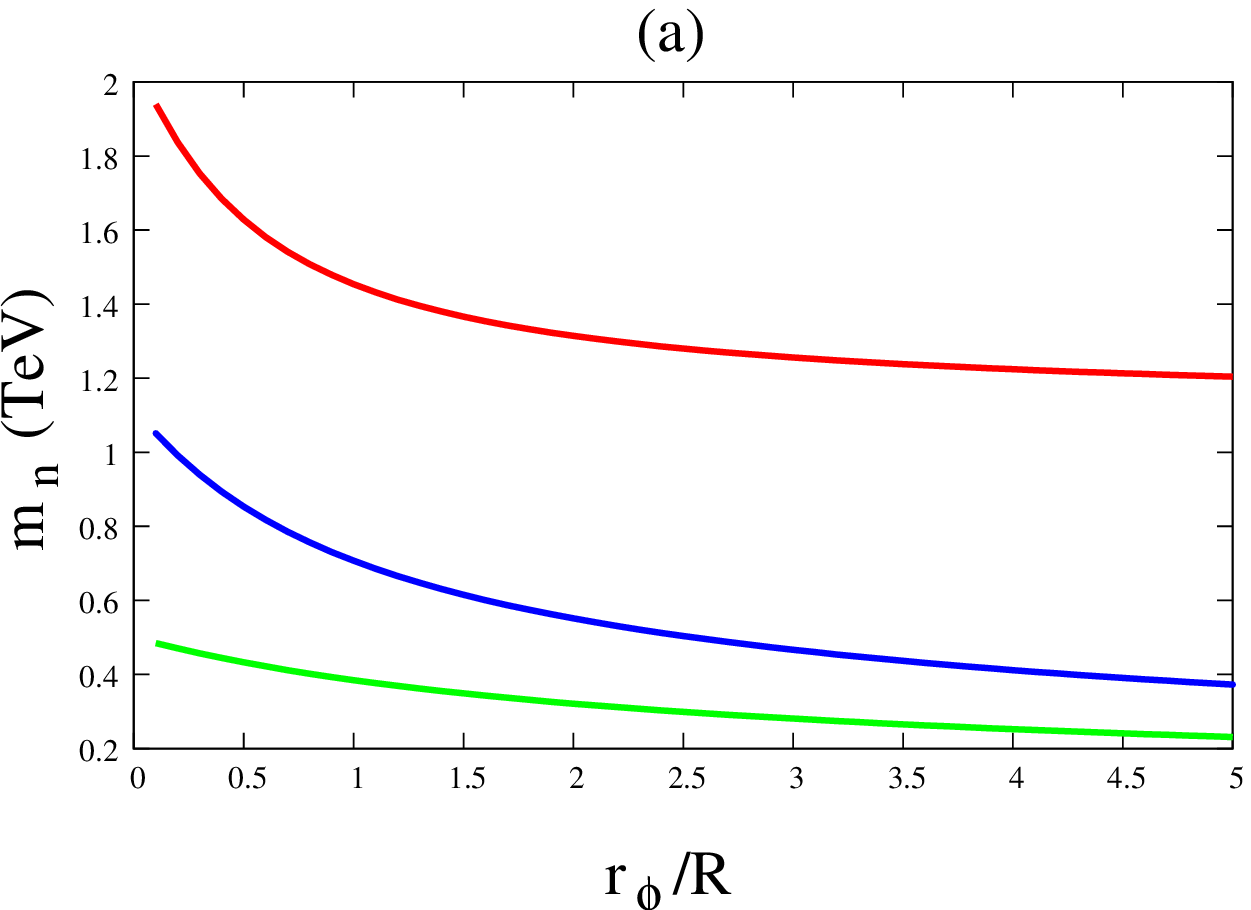}}
\resizebox{7.5cm}{!}{\includegraphics{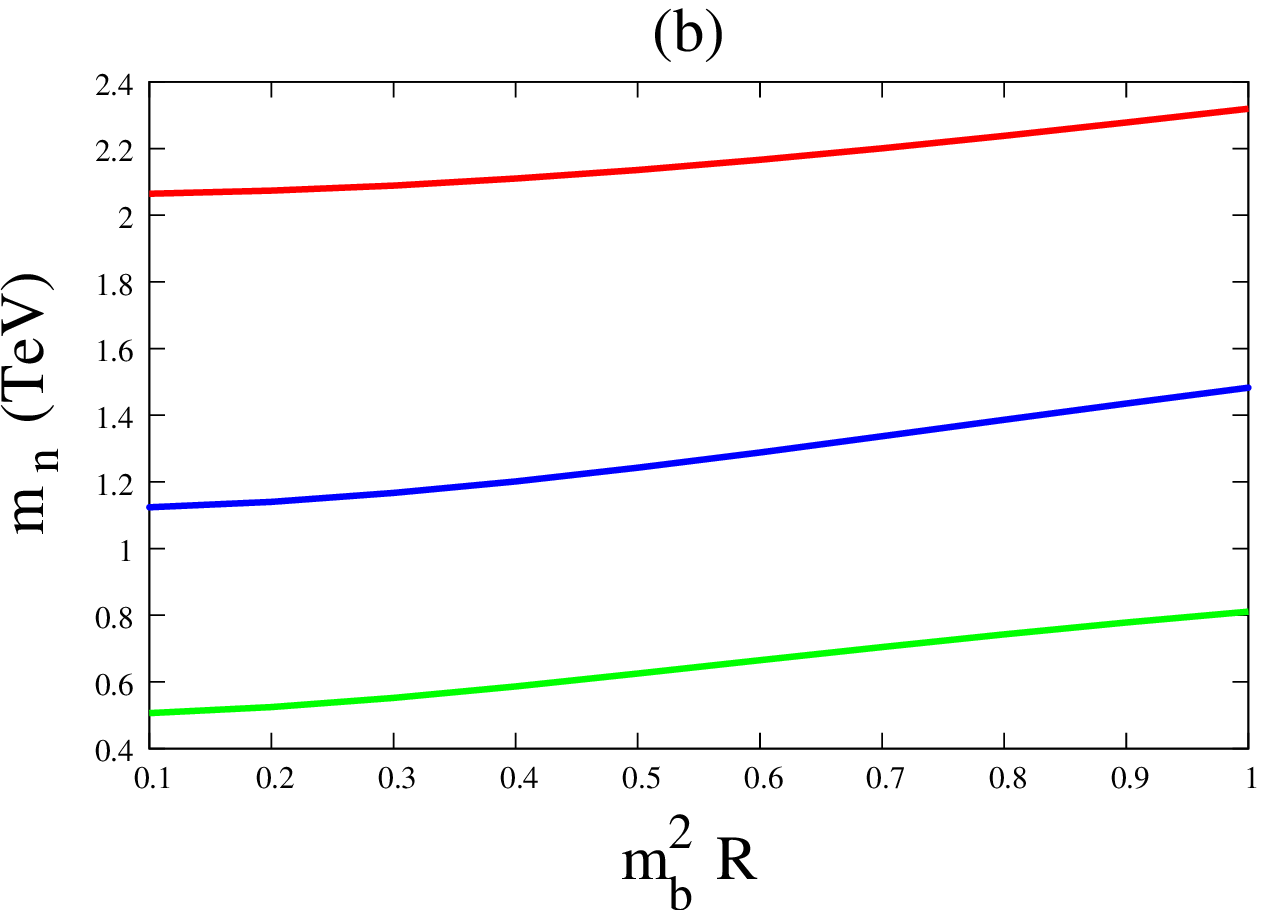}}
\end{center}
\caption{(a) Variation of the KK spectrum as a function $r_\Phi$ with
$R^{-1}=1$~TeV, $m=.5$~TeV $m_b=0$.
(b) Variation of the KK spectrum for as a function of $m_b$ with
$R^{-1}=1$~TeV, $m=.5$~TeV $r_\Phi=0$. The green (light gray) curve corresponds
to the variation of the zero mode mass, the blue (black) curve corresponds to
the variation of the first KK mode mass, and the red (dark gray) corresponds to
the variation of the second KK mode mass.}
\label{Figscalar}
\end{figure}

In Fig.~\ref{Figscalar} we
show the variation of the mass spectrum for different values of $r_\Phi$
and $m_b$. Boundary localized kinetic terms have the effect of decreasing the
n$^{\rm th}$ KK mass below that of $\sqrt{(n/R)^2 + m^2}$
while boundary localized mass terms have the opposite effect of
increasing KK masses. In the limit $r_\Phi/R\to \infty$, the
second KK mode mass, $m_2$, is bounded from below by $\sqrt{(1/R)^2+m^2}$,
while
zero and first KK modes become massless. Thus, when identifying the zero mode
with a Standard Model
particle, the mass splitting between the first and second KK mode can be made
arbitrarily large.

\section{UED with boundary localized terms}\label{secUEDtheory}

In this section we apply the results of the scalar toy model to the
full UED spectrum. We include only electroweak BLTs because
many potentially viable dark matter candidates are present in the first
KK level of the electroweak sector. Hence the BLTs we
consider are
\bea
S_{BLT}&=&\int d^5 x \left[\delta(y)+\delta(y-\pi R)
\right]\times\nonumber\\
&&\left(- \frac{r_B}{4\gy^2}B_{\mu\nu}B^{\mu\nu} -
\frac{r_W}{4\gw^2}W^a_{\mu\nu}W^{a\mu\nu}\right. \nonumber\\
&&\left. + r_H (D^\mu H)^\dagger D_\mu H +\mu_b^2 H^\dagger H - \lambda_b (H^\dagger H)^2 \right) \label{FullBLT}
\eea
where $r_B,r_W,r_H$ are constants, which from naive dimensional analysis have a
natural value of the order of $\frac{6\pi}{\Lambda}$, where $\Lambda$ is the
cutoff scale \cite{Carena:2002me}.

The BLTs in Eq.~(\ref{FullBLT}) respect the $SU(2)\times U(1)_Y$
gauge symmetry and Lorentz invariance on the brane. Their presence
breaks 5 dimensional translation invariance, which, however, is already broken
by the
presence of the branes. We will work in the limit of zero brane thickness, so
that brane terms containing $\p_5$ do not affect the KK
spectrum~\cite{Carena:2002me,delAguila:2003bh} and therefore the kinetic terms
we consider are parallel to the brane.
For the boundary Higgs potential we assume the fine tuned condition $\vh\equiv
\sqrt{\muh^2/\lambdah}=\sqrt{\mu_b^2/\lambda_b}$, which guarantees, that the
expansion of the Higgs around the VEV $\hat{v}(y)=\vh=const$ is consistent with
the bulk and boundary variations.\footnote{For studies of electroweak symmetry
breaking in the presence of a non-constant VEV but in absence of boundary
kinetic terms see Refs~\cite{varvev}.}

Following the spirit of the last section, we need to find the bulk equations of
motion for all fields in the electroweak sector and then determine the
KK decomposition from their boundary conditions. The boundary conditions
include
terms from the variation of the bulk action on the boundary as well as BLTs.
For the electroweak sector of UED, the situation is complicated by the mixing
between the $B_\mu$ and $W^3_\mu$ gauge bosons, as well as the mixing between
the $B_5,W^3_5$, and the Higgs field which contain the Goldstone and the
physical Higgs modes. In order
to identify the physical Higgs modes and the correct boundary conditions for
the Higgs and gauge fields, we reformulate the KK decomposition procedure of
Ref.~\cite{Muck:2001yv} along the lines of Ref.~\cite{Cacciapaglia:2005pa} in
Appendices~\ref{appgaugefix} and \ref{appgaugefixbd}. The manifestly 5
dimensional formulation of the Goldstone bosons and Higgs fields enables us to
incorporate the BLTs of Eq.~(\ref{FullBLT}) and determine the boundary
conditions.

In the following subsections, we present the bulk equations of motion (which are
unmodified by the boundary terms), their boundary conditions as derived in
Appendices~\ref{appgaugefix} and \ref{appgaugefixbd}, and discuss the mass
spectrum and couplings of all zero and KK modes in the electroweak sector. For
convenience, the results are summarized in Appendix~\ref{appSpecs}.

\subsection{The Higgs mass spectrum and wavefunctions}\label{secHiggsSpec}

For the Higgs, the bulk equations of motion and boundary conditions are given
by
\bea
\left[\Box-\p_5^2 + 2 \muh^2\right]h &=& 0\label{hEOM}\\
\left[\pm\p_5+(r_H \Box+2\mu_b^2)\right] h |_{y=\pi R,0}&=&0,\label{hbdvar}
\eea
where the $+\p_5$ ($-\p_5$) corresponds to the boundary condition at $\pi R$
$(0)$, which is similar to those of massive scalar in
Section~\ref{secmassiveBLKT}.
Hence we can determine the mass quantization conditions, wavefunctions and
normalization factors from the corresponding results of
Section~\ref{secmassiveBLKT} using the identifications: $r_\Phi \to r_H$,
$m^2_\Phi \to 2\muh^2$,  and $m^2_b \to 2\mu^2_b$.

The Higgs KK masses can be lowered by increasing $r_H$ and raised by increasing
$\mu_b$. As the Standard Model Higgs has not been found, the only
phenomenological bound on $\muh$ and $\mu_b$ is, that the zero mode mass must
be greater than $ 115$~GeV. However in the
next section we will see that $r_H$ is constrained because it influences
the gauge boson masses.

\subsection{Mass spectrum of the charged electroweak sector}\label{secChargedSpec}

The charged gauge bosons are decomposed so that their equations of motion
and boundary conditions are\footnote{The details of
gauge fixing in the presence of boundary terms is discussed in
Appendices~\ref{appgaugefix} and \ref{appgaugefixbd}. The results we quote here
are in unitary gauge.}
\bea
\left[\left(\Box-\p_5^2+\frac{\gw^2\vh^2}{4}\right)\eta^{\mu\nu}-\p^\mu \p^\nu
\right]W^\pm_\mu &=& 0 \label{WEOM}\\
\left(\pm\p_5+r_W\left[\Box \eta^{\mu\nu}-\p^\mu\p^\nu\right]+r_H\left(\frac{
\gw\vh}{2}\right)^2\right) W^\pm_ \nu |_{y=\pi R,0} &=&0\label{Wbdvar}
\eea
which resembles those of the scalar model in Section~\ref{secmassiveBLKT}.
Again, the mass determining equations and wavefunctions can be read off from
Section~\ref{secmassiveBLKT} using the identifications $r_\Phi \to r_W$,
$m^2_\Phi \to \gw^2 \vh^2/4$, and $m^2_b \to r_H \gw^2\vh^2/4$. For
the normalization conditions, we rescale Eq.~(\ref{normcond}) by $\gw^2$ so
that
\beq
\frac{1}{\gw^2}\int_{0}^{\pi R} dy \left[1+ r_W \left[\delta(y) + \delta(y-\pi
R) \right]
\right] f_i^W f_j^W = \delta_{ij},\label{normcondW}
\eeq
which guarantees the canonical normalization of kinetic terms for each KK mode.
The boundary ``mass'' term for the $W$ boson is induced by the
boundary interaction term $\mathcal{L}_{BLT}\supset \frac{r_H}{4} H^{\dagger}
H W^+_\mu
W^{-\mu}$ in Eq.~(\ref{FullBLT}) when electroweak symmetry breaking occurs.
Note that the condition for a flat zero mode in Eq.~(\ref{flatnesscondPhi})
translates into $r_W=r_H$. Unless the Higgs and gauge BLKT parameters
are identical, $W^{\pm(0)}$ has a $y$-dependent profile.

Apart from the charged gauge bosons, as explained in
Section~\ref{secUEDreview},
UED also contains charged Higgses at non-zero KK levels. According to
Appendices~\ref{appgaugefix} and~\ref{appgaugefixbd}, the equation of
motion and boundary conditions of the charged Higgs boson are
\bea
\left(\Box - \p_5^2 + \frac{\gw^2 \vh^2}{4} \right) a^{\pm} &=& 0
\label{apmEOM}\\
(a^\pm \pm r_H \p_5 a^\pm) |_{y=\pi R,0} &=& 0.
\eea
Using the results of Section~\ref{secmassiveBLKT} along with the identification
$r_\Phi \to r_H$, $m^2_\Phi \to \gw^2 \vh^2/4$, and $m^2_b \to r_H \gw^2\vh^2/4
$, we can find the $a^{\pm}$ mass determining equations and the wavefunctions.
The orbifold condition projects out the zero mode of the charged Higgs boson
and the higher KK mode wavefunctions are purely ``trigonometric'' of the form in
Eq.~(\ref{phibasis}).

\subsection{Mass spectrum of the neutral electroweak sector}

In the neutral sector, the KK decomposition is complicated by the fact that
$B_\mu$ and $W^3_\mu$ mix in the bulk as well as on the boundary. In the
special case $r_B=r_W$, the bulk and the boundary action can be diagonalized by
the same 5 dimensional field redefinition. We present this case first and
indicate the generalization to $r_B\neq r_W$ in Section~\ref{secBneqW}.

\subsubsection{The special case $r_B=r_W\equiv r_g$}\label{secrBeqrWth}

After electroweak symmetry breaking the 5 dimensional mass matrix in
Eq.~(\ref{UEDbulkaction}) is diagonalized by the 5 dimensional field
redefinition
\bea
Z_M&=&\frac{1}{\sqrt{\gy^2+\gw^2}}\left(W^3_M-B_M\right)\nonumber\\
A_M&=&\frac{1}{\sqrt{\gy^2+\gw^2}}\left(\frac{\gy}{\gw}W^3_M+\frac{\gw}{\gy}B_M
\right).
\label{ZAbulkbasis}
\eea
If $r_B=r_W\equiv r_g$, this simultaneously diagonalizes the boundary mass terms
induced by electroweak symmetry breaking in Eq.~(\ref{FullBLT}). The equation
of motion and boundary conditions for the $Z_{\mu}$ gauge
field are then
\bea
\left[\left(\Box-\p_5^2+\frac{(\gy^2+\gw^2)\vh^2}{4}\right)\eta^{\mu\nu}-\p^\mu
\p^\nu \right]Z_\mu &=& 0 \label{ZEOM}\\
\left(\pm\p_5+r_g\left[\Box \eta^{\mu\nu}-\p^\mu\p^\nu\right]+r_H\frac{(\gy^2+
\gw^2)\vh^2}{4}\right) Z_\mu|_{y=\pi R,0} &=&0\label{ZBbdvar}
\eea
while those for $A_{\mu}$ are
\bea
\left[\left(\Box-\p_5^2\right)\eta^{\mu\nu}-\p^\mu \p^\nu \right]A_\mu &=& 0
\label{AEOM}\\
\left(\pm\p_5+r_g\left[\Box \eta^{\mu\nu}-\p^\mu\p^\nu\right]\right) A_\mu
|_{y=\pi R,0} &=&0.\label{ABbdvar}
\eea
Similar to the charged gauge bosons the mass determining relations and
wavefunctions are found using the results of Section~\ref{secmassiveBLKT}
with the identifications $r_\Phi \to r_g$, $m^2_\Phi \to (\gw^2 +\gy^2)
\vh^2/4$, and $m^2_b \to r_H (\gw^2 + \gy^2)\vh^2/4$ for $Z_\mu$ and $r_\Phi
\to r_g$, $m^2_\Phi \to 0$, and $m^2_b \to 0$ for $A_\mu$.
Hence the zero mode of the photon is always flat, while the zero mode of the
$Z$ will be flat only when $r_H=r_g$. The normalization of the $Z$ and
photon is the same as in Eq.~(\ref{normcondW}) with replacement $\gw^2 \to 1$
due to the basis chosen in Eq.~(\ref{ZAbulkbasis}).

Similar to the charged bosons in Section~\ref{secChargedSpec},
the neutral sector contains a physical pseudoscalar degree of freedom. Its
equation of motion and boundary conditions are
\bea
\left(\Box - \p_5^2 + \frac{(\gy^2+\gw^2) \vh^2}{4} \right) a^0 &=& 0
\label{apmEOMrwrb}\\
(a^0 \pm r_H \p_5 a^0) |_{y=\pi R,0} &=& 0.
\eea
The wavefunctions and mass determining equations follow
directly from the charged Higgs case discussed in the last section with the
replacement $\gw^2\to (\gy^2+\gw^2)$.

\subsubsection{The general case: $r_B\neq r_W$}\label{secBneqW}

If $r_B\neq r_W$, the field redefinition in Eq.~(\ref{ZAbulkbasis}) no longer
decouples the boundary conditions following from Eq.~(\ref{FullBLT}) and, at
least for the neutral gauge fields, we have to refine the strategy to find the
KK decomposition. Before doing so note that the boundary parameters $r_B$ and
$r_W$ in the boundary action Eq.~(\ref{ZAbulkbasis}) only affect the boundary
conditions of the gauge fields. The KK decompositions of the physical Higgs
bosons are not affected and therefore remain the same as in the $r_B=r_W$
case discussed in the last section.

Aside from the mixing of $B_\mu-W^3_\mu$, the KK decomposition for $W^3_\mu$ is
identical
to those of $W^\pm$ discussed in Section~\ref{secrBeqrWth}. The analogous
solutions for the $B_\mu$ are also given by the substitutions $r_W\to r_B$ and
$\gw \to \gy$ for the relations in Section~\ref{secrBeqrWth}. The mixing
term in the bulk $\propto \hat{v}^2 B_\mu W^\mu$ and on the brane $\propto r_H
\hat{v}^2 B_\mu W^\mu$ induce off-diagonal terms in the neutral gauge boson
mass matrix of the form
\bea
\mathcal{M}_{m,n}^2 &=& \frac{\hat{v}^2}{4} \int_0^{\pi R} dy f_n^W(y) f_m^B(y)[1+r_H(\delta(y)+\delta(y-\pi R))]
\label{wbmixterm}
\eea
where $f_n^W(y)$ and $f_m^B(y)$ are the $W_{\mu}^3$ and $B_{\mu}$ wavefunctions
respectively. The even (odd) modes of $B_\mu$ only have a non-zero overlap with
the even (odd) modes of $W^3_\mu$, thereby still preserving KK parity. In
particular there is a non-trivial overlap between the zero mode of the $B_\mu$
gauge boson and all even modes of the $W^3_{\mu}$ gauge boson. In order
to find the exact mass spectrum and wavefunctions of the neutral sector, the
full KK mass matrix in the neutral sector has to be diagonalized.

\subsection{Modifications of zero mode and KK mode couplings}\label{secCouplMod}

Using the wavefunctions and mass spectra for all electroweak fields derived in
the last section, we can calculate the couplings of all KK particles by
integrating out the extra dimension.

The coupling of fermions to the $W$ or $B$ bosons arise from the 5 dimensional
action
\beq
S=\int d^5x \bar{f}\gamma^MD_M f
\eeq
where $D_M$ is appropriate covariant derivative. Assuming that fermions have
no BLTs, the couplings are
\beq
 g^f_{i,lmn}=\int dy f^{W,B}_l f^{f}_m f^{f}_n \label{gw4Dto5D}
\eeq
where $i=B,W$ and $\{f_m^f\}$ is the wavefunction of the $m^{th}$ KK mode. In
particular, the couplings of the zero modes are given by
\beq
 g^f_{i,000}=\frac{1}{\pi R}\int dy f^{W,B}_0 \label{gw4Dto5D0},
\eeq
where we used that the fermion zero modes are constant in the absence of fermion BLTs.

Now, let us consider the triple and quartic gauge boson vertices of the $SU(2)$. They follow from the overlap integrals
\bea
g^t_{2,lmn}&=&\int dy f^{W}_l f^{W}_m f^{W}_n\left(1+r_W\left[\delta(y)+\delta(y-\pi R)]\right]\right)\label{gw4Dto5Dtr}\\
\left(g^q_{2,klmn}\right)^2&=&\int dy  f^{W}_k f^{W}_l f^{W}_m f^{W}_n\left(1+r_W\left[\delta(y)+\delta(y-\pi R)]\right]\right)\label{gw4Dto5Dqu}.
\eea
For generic values of $r_W$ it is obvious that $g^f_{i,000} \neq g^t_{2,000}
\neq g^q_{2,0000}$. As the zero mode level is identified with the Standard
Model, the BLKTs induce a modification to the $WWZ$ vertex.

The vertices of the Higgs KK modes with gauge bosons and fermions are
calculated analogously from overlap integrals, taking the BLTs into account as
in  Eq.~(\ref{gw4Dto5Dtr}) and Eq.(\ref{gw4Dto5Dqu}). Using the mass
determining equations and wavefunctions of the previous subsections, for fixed
$(R,r_B,r_W,r_H)$ we can match the zero mode spectrum and couplings to
those
of the Standard Model and hence determine the spectrum and coupling of the
higher KK modes.

\section{Phenomenology of Electroweak BLTs in UED}\label{secPheno}

As shown in the last section, BLTs can significantly modify the KK spectrum and
gauge couplings in UED. The modification of the spectrum can alter the nature
of the lightest KK particle so that $B^{(1)}$ is no longer the LKP. However
these BLTs can also affect the zero mode couplings and masses and therefore are 
constrained by experiment.

In this section we study the constraints on $r_B,r_W$, and $r_H$ arising from
matching the zero mode spectrum of non-minimal UED to that of the Standard
Model. We adopt the strategy of fixing the parameters $(r_B,r_W,r_H,R)$
and solving for the 5 dimensional quantities $(\gy,\gw,\vh)$ in terms of
the Standard Model observables $\alpha_{em},G_F,m_Z,m_W$. Therefore the
5 dimensional quantities $(\gy,\gw,\vh)$ are over constrained and provide
a bound on the parameters $(r_B,r_W,r_H,R)$. A further potential bound arises
from the LEP measurement of the $WWZ$ vertex.

In this article we discuss three qualitatively differing regions of the
non-minimal UED parameter space. The first parametric scenario is one in which
all the electroweak boundary localized kinetic terms are uniform and all other
Higgs boundary terms are zero. In this scenario
the LKP is the KK photon where the Weinberg angle is the same at every KK
level. However as KK number is explicitly broken by the BLKTs, the even KK
modes of $W^{\pm}$ and $Z$ have non-zero couplings to the zero mode fermions
which leads to non-trivial contributions to the Fermi constant $G_f$.

The second parametric scenario we consider in Section~\ref{secPhenoBeqW} is the
case $r_g\equiv r_W=r_B \neq r_H$, with all other Higgs BLTs zero. As $r_H$ is
split from $r_g$, the zero
mode wavefunctions of the gauge bosons are not flat which leads to modified
gauge couplings. We calculate the bounds on $r_H-r_g$ arising from modifications of
the zero mode mass spectrum. In this scenario we find regions of parameter space
in which $h^{(1)}$ is the LKP.

The third scenario we consider in Section~\ref{secPhenoBneqW}, is the
case in which $r_W \neq r_B=0=r_H$ and all the other Higgs brane terms are
zero. As $r_W \neq r_B$, the bulk and brane mass terms of the neutral gauge boson
sector induce mixing between different KK levels of the $B$ gauge boson and the
the $W^3$ gauge boson. For large enough $r_W$ the
$W^{3(1)}$ becomes lighter than the $B^{(1)}$, in which case the LKP becomes
mainly $W^{3(1)}$. As for the previous scenarios, we determine bounds on $r_W$ which arise
due to the modified zero mode couplings. We show that within these bounds, a $W^{3(1)}$
LKP can easily be realized.

\subsection{Scenario I: $r_W=r_B=r_H$}\label{secPhenoUnivr}

\begin{figure}
\begin{center}
\resizebox{7.5cm}{!}{\includegraphics{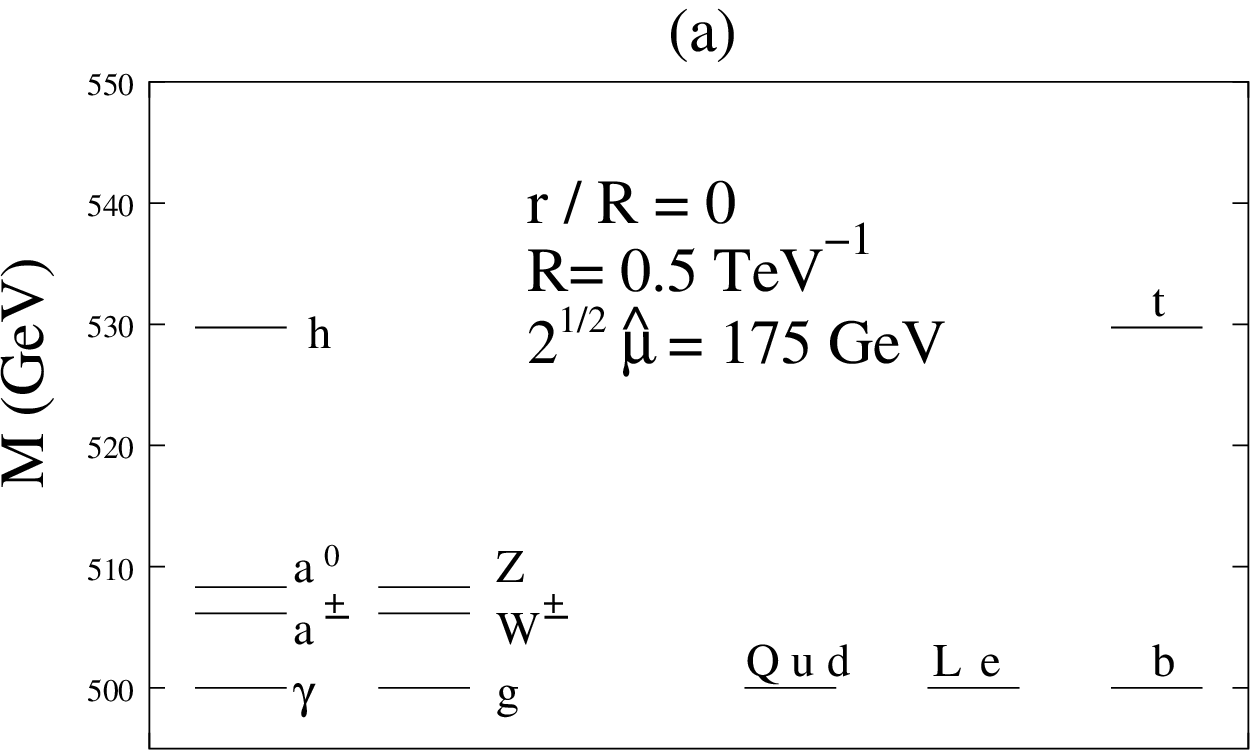}}
\resizebox{7.5cm}{!}{\includegraphics{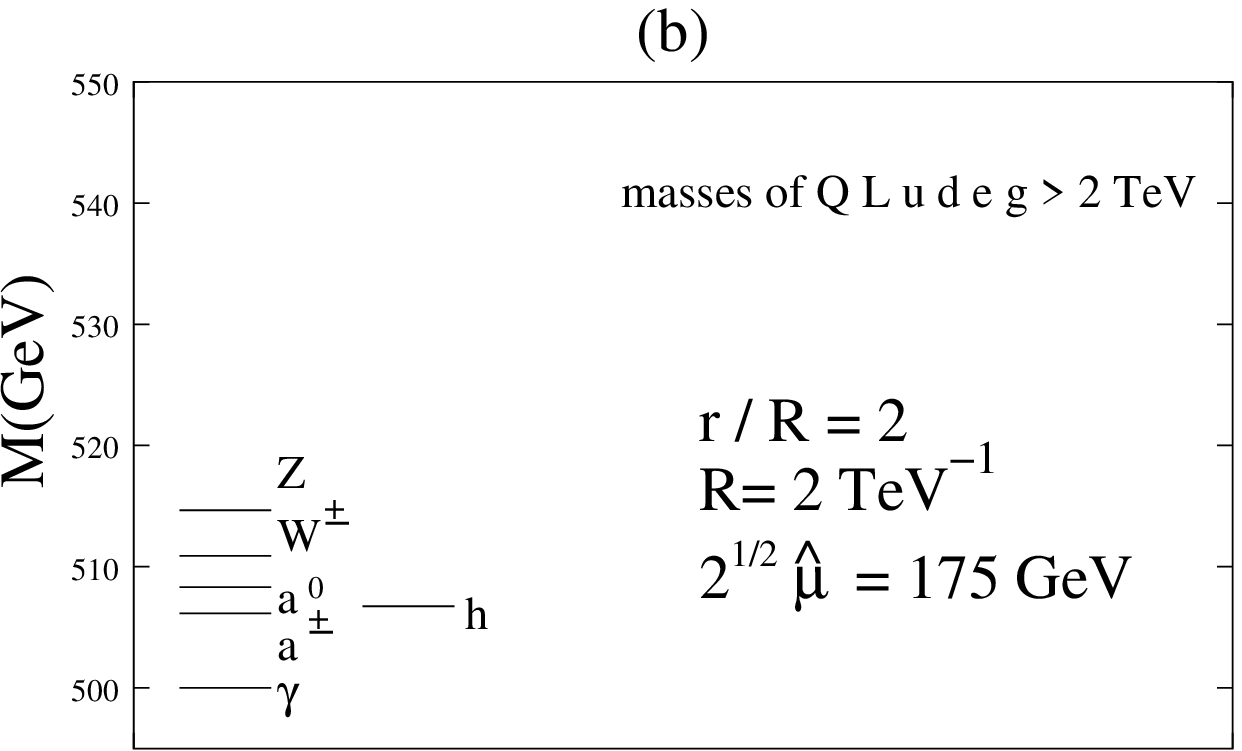}}
\end{center}
\caption{Sample spectra for UED.  (a) The tree level UED spectrum without
BLTs (first KK level) for $R^{-1}=500$ GeV. (b) The tree level UED spectrum
with $r_B= r_W=r_H=2 R$, $\mu_b=0$, $R^{-1}=1$~TeV and $m_h=115$~GeV.
Parameters are chosen such that the LKP masses of both model
coincide.}
\label{fig2}
\end{figure}

For a start, let us consider the special case of uniform gauge BLKTs $r_{EW}\equiv r_W=r_B=
r_H$. Assuming
the absence of boundary terms for the fermions and for the gluon, this model
presents a simple extension of UED with only four free parameters $R,r_{EW},
\mu_b,m_h$. For simplicity, we assume $\mu_b=0$ and $m_h=115$~GeV.\footnote{Relaxing these assumption results in heavier Higgs KK masses, but otherwise does not affect the mass spectrum.}
For uniform $r_{EW}$, the boundary conditions on the gauge fields imply that
all gauge field zero modes are flat. Therefore the matching of the underlying 5
dimensional parameters $\gy,\gw,\vh$ to the Standard Model values $g_Y,g_2,v$
is similar to ``standard'' UED, with $\pi R \to \pi R + 2 r_{EW}$ in the
rescaling of the couplings, but the identification of $\muh$ with
$m_h$ is altered. At the first KK level, the
fermions and gluon have a mass $\sim n/R$ but the gauge bosons and Higgs masses
are reduced to lower values. Thus $R^{-1}$ determines the fermion and gluon mass
scale, while $r_{EW}$ can be thought of as parameterizing the mass splitting
between the electroweak KK modes and the fermion and gluon KK modes. In
Fig.~\ref{fig2} we show the effect of $r_{EW}$ on the particle
spectrum of UED. Fig.~\ref{fig2}(a) corresponds to ``standard'' UED
with $R^{-1}=0.5$~TeV while  Fig.~\ref{fig2}(b) corresponds to the
uniform $r_{EW}$
scenario with $\mu_b=0, r_{EW}/R =2$ and $R^{-1}=1$~TeV. In spite of
different values of $R$ it is possible to choose $r_{EW}/R$, such that the
LKP mass is the same in both scenarios.

From the equations of motion and boundary conditions in
Sections~\ref{secHiggsSpec},~\ref{secChargedSpec} and~\ref{secrBeqrWth} we can
see that for uniform BLKTs the spectrum has the following structure
\bea
m_{Z^{(1)}}&\geq&m_{a^{0(1)}}>m_{a^{\pm(1)}}>m_{\gamma^{(1)}}\\
m_{Z^{(1)}}&>&m_{W^{\pm(1)}}\geq m_{a^{\pm(1)}}\\
m_h^{(1)}&>&m_{\gamma^{(1)}}.
\eea
Hence, the LKP is always the KK photon or, more precisely, the linear
combination $\gamma^{(1)}=\sin(\theta_W^{(1)})B^{(1)}+\cos(\theta_W^{(1)})W^{3(1)}
$, where for uniform $r_{EW}$ the Weinberg angle $\theta^{(n)}_W$ at all KK levels is identical to that of the Standard Model.

As the BLTs do not introduce any extra sources flavor
violation and the fermion KK modes are heavier than in ``standard'' UED,
the UED GIM mechanism~\cite{Buras:2002ej} implies weaker flavor constraints.
As the Weinberg angle is the same at every KK level, the LKP can annihilate
efficiently through a t-channel $W^{(1)}$ into $W^+ W^-$ even if the KK
fermions
are quite heavy. In the limit of very heavy KK fermions, the requirement that
the KK photon does not overclose the universe implies an upper bound on LKP
mass of about $1.6$~TeV. This bound is a constraint on the LKP mass which can
be substantially smaller than the compactification scale $R^{-1}$.

The collider constraints and electroweak constraints can be strong in this
scenario. As KK number is explicitly broken by the BLTs, higher KK level gauge
bosons have non-zero couplings to the zero mode fermions. Hence there are
 electroweak corrections even at tree level. For example, the Fermi constant
$G_f$ obtains contributions from the exchange of all even $W^{\pm(n)}$ KK
modes. In the next section, we study constraints arising from the tree level
modifications of the zero mode and KK mode couplings in detail.

\subsubsection{Electroweak constraints} \label{rewcon}

In order to determine constraints from the electroweak sector we first match the zero mode spectrum on to
the Standard Model, using the results of
Sections~\ref{secHiggsSpec},~\ref{secChargedSpec} and~\ref{secrBeqrWth}.
We perform the matching by demanding that $\gw, \gy$ and $\vh$ are chosen
so that $\alpha,G_f,m_W$ and $m_Z$ have their correct values within
experimental errors, for fixed $r_{EW}$ and $R$. As there are three underlying parameters
and four Standard Model quantities, the system is over constrained.
Hence we can predict one of the Standard Model parameters which we use to
constrain the input parameters $r_{EW}$ and $R$.

For uniform $r_{EW}$ the gauge boson zero modes are flat and their masses are directly related to the 5
dimensional parameters by
\bea
m_W^2 &=& \mwh^2 \equiv \frac{\gw^2 \vh^2}{4} \label{mWreln}\\
m_Z^2 &=& \mzh^2 \equiv \frac{(\gw^2 + \gy^2) \vh^2}{4},
\eea
giving us two relations between the 4 dimensional and 5 dimensional parameters. With $\mwh$ and $\mzh$ determined, all $W^{\pm}$ and $Z$ KK mode wavefunctions and masses are can be found by numerically solving the mass determining equations given in Sections~\ref{secHiggsSpec},~\ref{secChargedSpec} and~\ref{secrBeqrWth} which are summarized in Eq.~(\ref{nQcondHBLKT}) in Appendix~\ref{appSpecs}.

The Fermi constant $G_f$ gets contributions from the exchange of the $W^{\pm(0)}$ mode as well as higher even KK modes. Summing over all $W$ KK modes that have non-zero couplings to the zero mode fermions we find the effective Fermi constant
\beq
G_f=\frac{\gw^2}{4 \sqrt{2} \pi R} \sum_{n=0}^{\infty} \frac{b_{2n}}{
m^2_{W^{(2n)}}} \label{GfdefrH}
\eeq
where
\beq
b_{0} = \frac{1}{1+\frac{2 r_{EW}}{\pi R}},
\eeq
\beq
b_{2n} = \left(\frac{8\sin^2 \frac{M_{2n}^W \pi R}{2}}{
\left(1 + \frac{\sin M_{2n}^W \pi R}{M_{2n}^W \pi R} +\frac{4 r_{EW}}{\pi R}
 \cos^2 \frac{M_{2n}^W \pi R}{2}\right) \left(M_{2n}^W \pi
R \right)^2}\right), \label{bcoeffdef}
\eeq
$M_{2n}^W=m_{W^{\pm(2n)}}^2-\mwh^2$, and $m_{W^{\pm(2n)}}$ are the physical masses of the $W^\pm$ KK modes.

The $U(1)_{em}$ coupling in terms of the the 5 dimensional couplings is
\beq
\alpha_{em}=\frac{1}{4\pi(\pi R+2r_{EW})}\frac{\gy^2\gw^2}{\gy^2+\gw^2}.
\label{alphadef_rew}
\eeq
where the factor $\pi R+2r_{EW}$ comes from the normalization of the gauge
boson zero modes with BLKTs $r_{EW}$.

As we are studying the tree level corrections to the Standard Model relations,
we fix the values of $\alpha = 1/(127.925 \pm 0.016)$, $m_W =
(80.398 \pm 0.025)$~GeV and $G_f = (11.66367 \pm 0.00005)$~TeV$^{-2}
$~\cite{PDG} and use Eq.~(\ref{mWreln}), Eq.~(\ref{GfdefrH}), and Eq.~(\ref{alphadef_rew})
 to predict the value of $m_Z^{nUED}(R,r_{EW},m_W,G_f,
\alpha)$ and compare it to
\bea
\lim_{R,r_{EW} \to 0} m_Z^{nUED}(R,r_{EW},m_W,G_f,\alpha) &\equiv&
m_Z^{tree}(m_W,G_f,\alpha) \nonumber \\
&=& m_Z^{exp}- m_Z^{SM}|_{loop}(m_W,G_f,\alpha),
\label{mztreedef}
\eea
where $m_Z^{exp}$ is the experimentally measured $Z$ mass while
$m_Z^{SM}|_{loop}(m_W,G_f,\alpha)$ is the Standard Model loop contribution to
the $Z$ mass. Therefore we can
translate all experimental uncertainties into a band of allowed values of
$m_Z^{tree}(m_W,G_f,\alpha)$ and compare it to the predicted band of values
for the tree level nUED $Z$ mass $m_Z^{nUED}(R,r_{EW},m_W,G_f,\alpha)$.

\begin{figure}
\begin{center}
\resizebox{7.5cm}{!}{\includegraphics{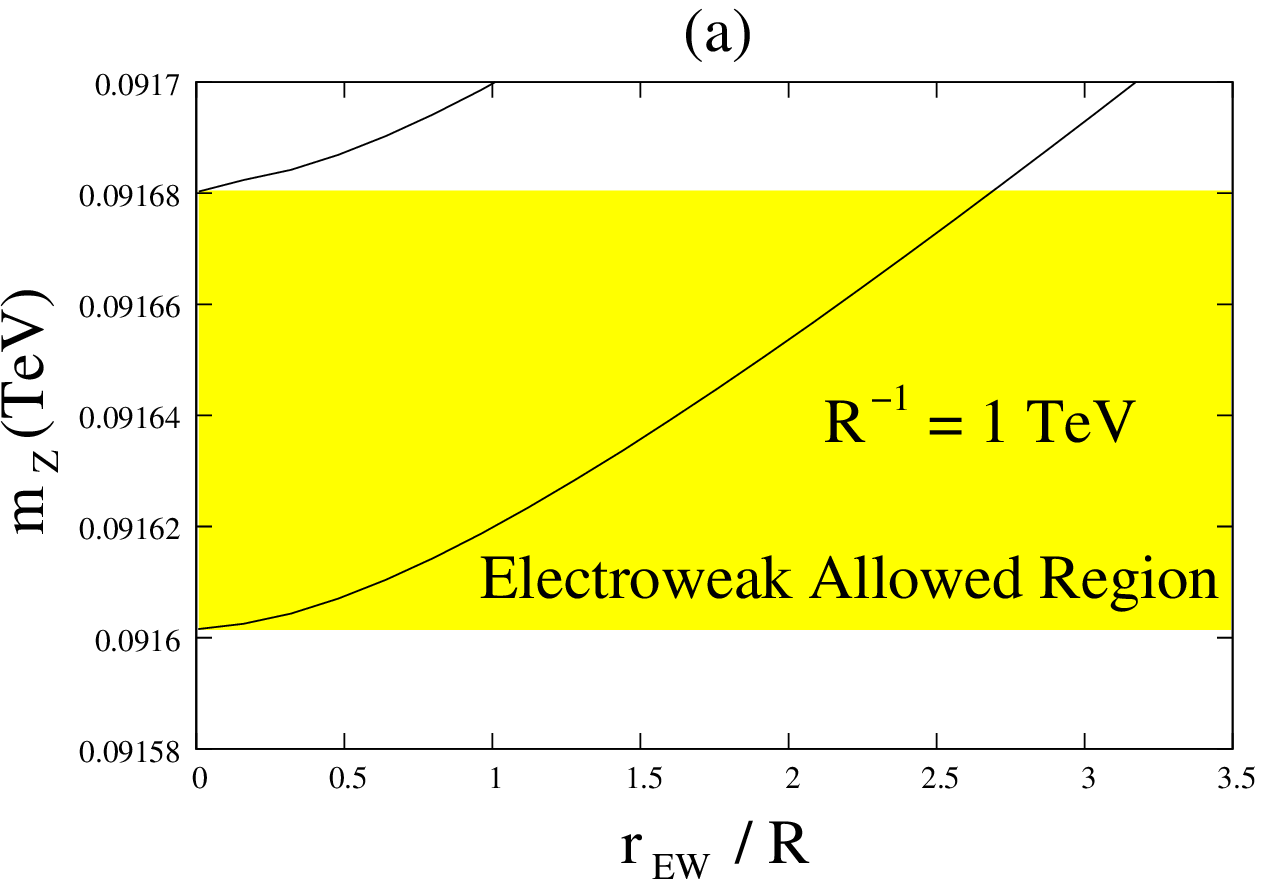}}
\resizebox{7.5cm}{!}{\includegraphics{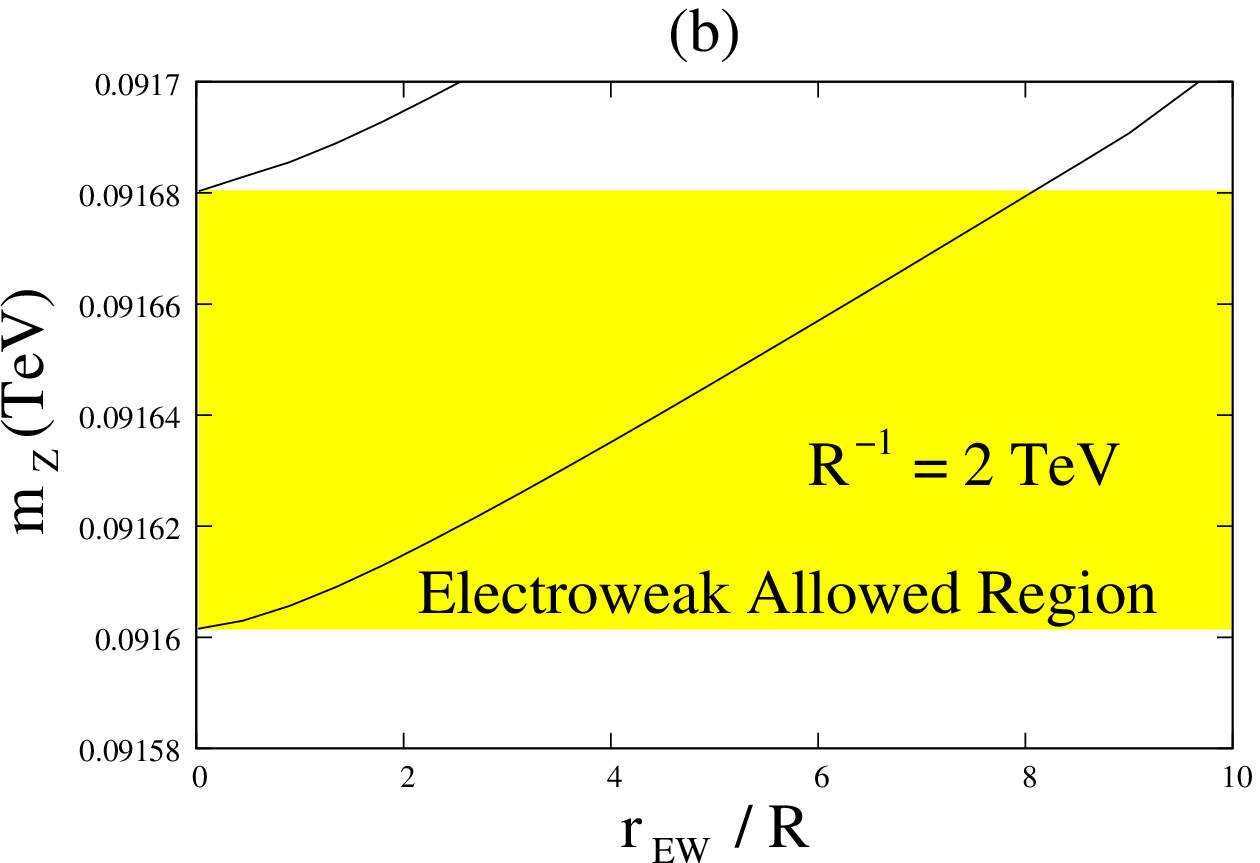}}
\end{center}
\caption{Variation of $m_Z^{nUED}$ for different values of $r_{EW}$
with $R^{-1}=1$~TeV and $R^{-1}=2$~TeV. The yellow (gray) band
corresponds to the 2$\sigma$ allowed tree-level value of $m_Z^{tree}$.
The region within the black lines is the 2$\sigma$ predicted tree-level value
of $m_Z^{nUED}$.}
\label{Figrew}
\end{figure}

Fig.~\ref{Figrew} presents the effect of varying $r_{EW}$
on the predicted value of $m_Z^{nUED}$ for $R^{-1}=1$~TeV
and $R^{-1} = 2$~TeV, assuming a 2$\sigma$ error in the input values of
$\alpha$, $m_W$ and $G_f$.
From Fig.~\ref{Figrew} we see that the constraints on $r_{EW}$ are weaker
for increasing $R^{-1}$. For $R^{-1} = 1$~TeV the electroweak constraints force $r_{EW}/R < 2.7$ while for $R^{-1} = 2$~TeV the bound only lies at $r_{EW}/R <8.0$. Thus for large compactification scales we can split
the fermion KK modes (with $m_f^{(1)}\sim 1/R$) from the gauge and Higgs KK
modes. Also note, that a higher compactification scale does \emph{not}
necessarily imply a heavier $\gamma^{(1)}$ LKP as can be seen from the sample
spectrum in Fig.~\ref{fig2}.

\subsection{Scenario II: $r_W=r_B\equiv r_g < r_H$}\label{secPhenoBeqW}

For uniform $r_{EW}$, the KK photon always remains the LKP, so to change the nature of the LKP we need
to split the BLKTs. As a next step let us consider a scenario in which we vary
the Higgs BLKT while keeping the gauge BLKTs equal at $r_B=r_W\equiv r_g$. For
$r_g>r_H$, it is obvious that the LKP remains the KK photon as the gauge boson
KK mode masses are reduced relative to the Higgs. We therefore focus on the choice $r_g<r_H$ for which it is conceivable that the $h^{(1)}$ becomes lighter than the KK photon.

For $r_H\neq r_g$, the gauge boson zero modes are not flat.
We thus need to find the correct values of $\gy,\gw$ and $\vh$ so as to match
the zero mode spectrum to that of the Standard Model. As in
Section~\ref{rewcon}, we perform the matching by demanding the correct values
for $\alpha,G_f,m_W$ and determine the tree-level value for $m_Z^{nUED}$, which
leads to a bound on $r_g$ and $r_H$.  In Sec.~\ref{SecrHLEPconstraints} we derive the
relations between the Standard Model parameters $\alpha,G_f,m_W,m_Z$ and the
underlying parameters  $(\gy,\gw,\vh)$ to study the constraints on
$(r_g,r_H,R)$. In Sec~\ref{rHDM}, we use this information in order to determine
the LKP in this scenario and show that there exist regions of parameter space
where $h^{(1)}$ is the LKP.

\subsubsection{Electroweak constraints}\label{SecrHLEPconstraints}

We work in the $A-Z$ basis defined in Eq.~(\ref{ZAbulkbasis}). From the appropriate
substitutions in Eq.~(\ref{nQcondHBLKT}) for $W^\pm$ and $Z$, the physical masses $m_{I^{(n)}}$ satisfy the
condition\footnote{For our parameter choice of $r_H>r_g$ the hyperbolic mass conditions in Eq.~(\ref{nQcondHBLKT})
do not have a solution, such that the zero mode in this case is given by a
cosine solution as well.}
\beq
\frac{r_g(M_n^I)^2-(r_H-r_g) \hat{m}_I^2}{M_n^I} = -\tan \frac{M_n^I
\pi R}{2} \label{WZquantrH}
\eeq
where $I = (W, Z)$, $\hat{m}_Z^2 = (\gw^2 + \gy^2)\vh^2/4$, $\hat{m}_W^2 =
\gw^2 \vh^2/4$ and $M_n^I = \sqrt{\hat{m}_I^2 - m_{I^{(n)}}^2}$.

As in the uniform $r_{EW}$ scenario, the effective Fermi constant obtains contributions from the exchange of all even $W^{\pm}$
KK modes as given earlier in Eq.~(\ref{GfdefrH}) with the non-zero KK coefficients $b_{2n}$ as given in Eq.~(\ref{bcoeffdef}) with $r_{EW}\rightarrow r_g$. Due to the non-flat zero mode profile, Eq.~(\ref{bcoeffdef}) also holds for the zero mode contribution $b_{0}$ in this scenario. Furthermore, working in the basis of Eq.~(\ref{ZAbulkbasis}), it is obvious that the
boundary mass term of the photon vanishes. Hence the relation between the
$U(1)_{em}$ coupling and the 5 dimensional couplings is again given by Eq.~(\ref{alphadef_rew}) with $r_{EW}\rightarrow r_g$.

To determine the allowed parameter space in $(r_H,r_g,R)$ we calculate
$\hat{m}_W$ by fixing the zero mode $W$ mass and using Eq.~(\ref{WZquantrH}).
Once we have found $\hat{m}_W$ we are able to determine all $W$ KK masses
and use Eq.~(\ref{GfdefrH}) to determine $\gw$. $\vh$ is determined via the
definition of $\hat{m}_W\equiv \gw \vh/2$.  Finally, using Eq.~(\ref{alphadef_rew}) we can
determine $\gy$ in terms of $\gw$. $m_Z$ is fixed by the
parameter set $(r_g,r_H,R,\gy,\gw,\vh)$ and a comparison with the experimental
values yields a constraint on $(r_g,r_H,R)$.
\begin{figure}
\begin{center}
\resizebox{7.5cm}{!}{\includegraphics{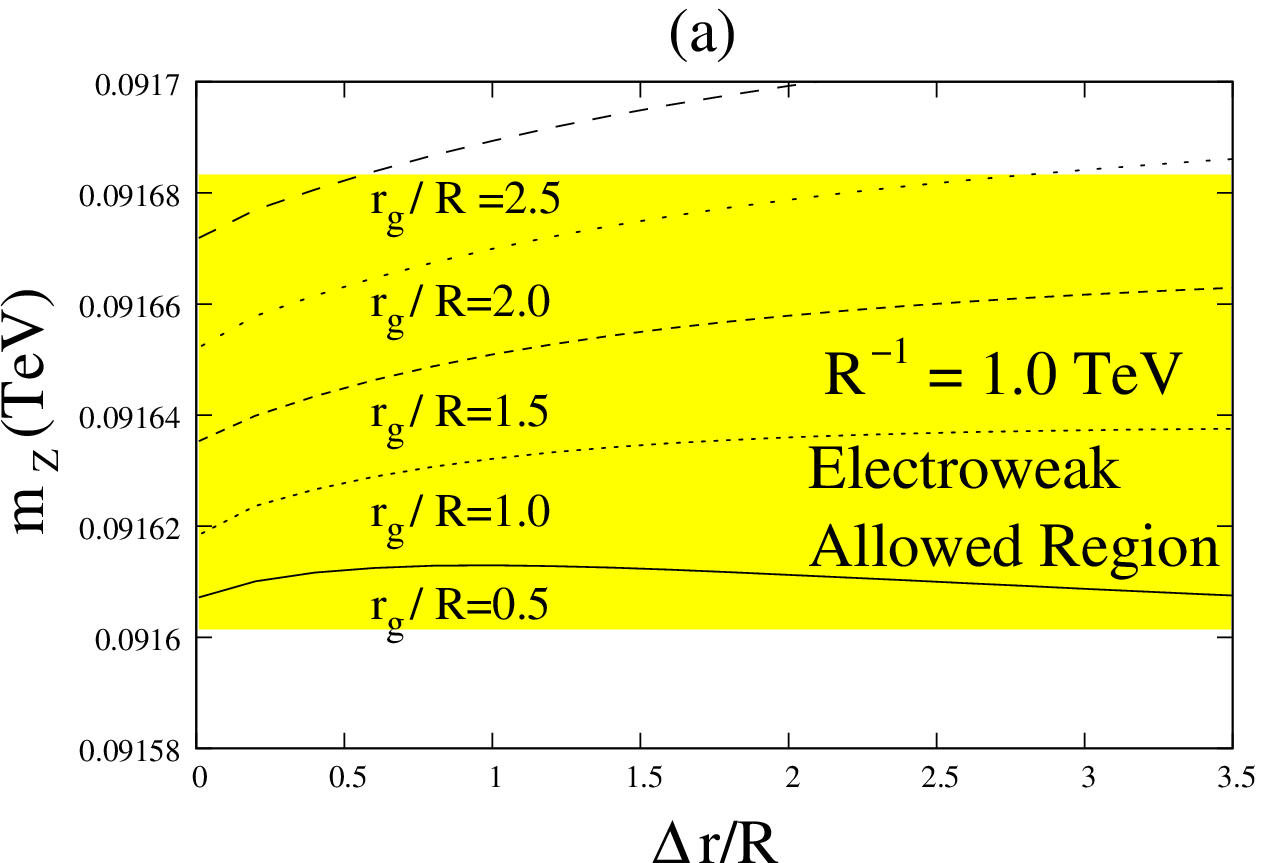}}
\resizebox{7.5cm}{!}{\includegraphics{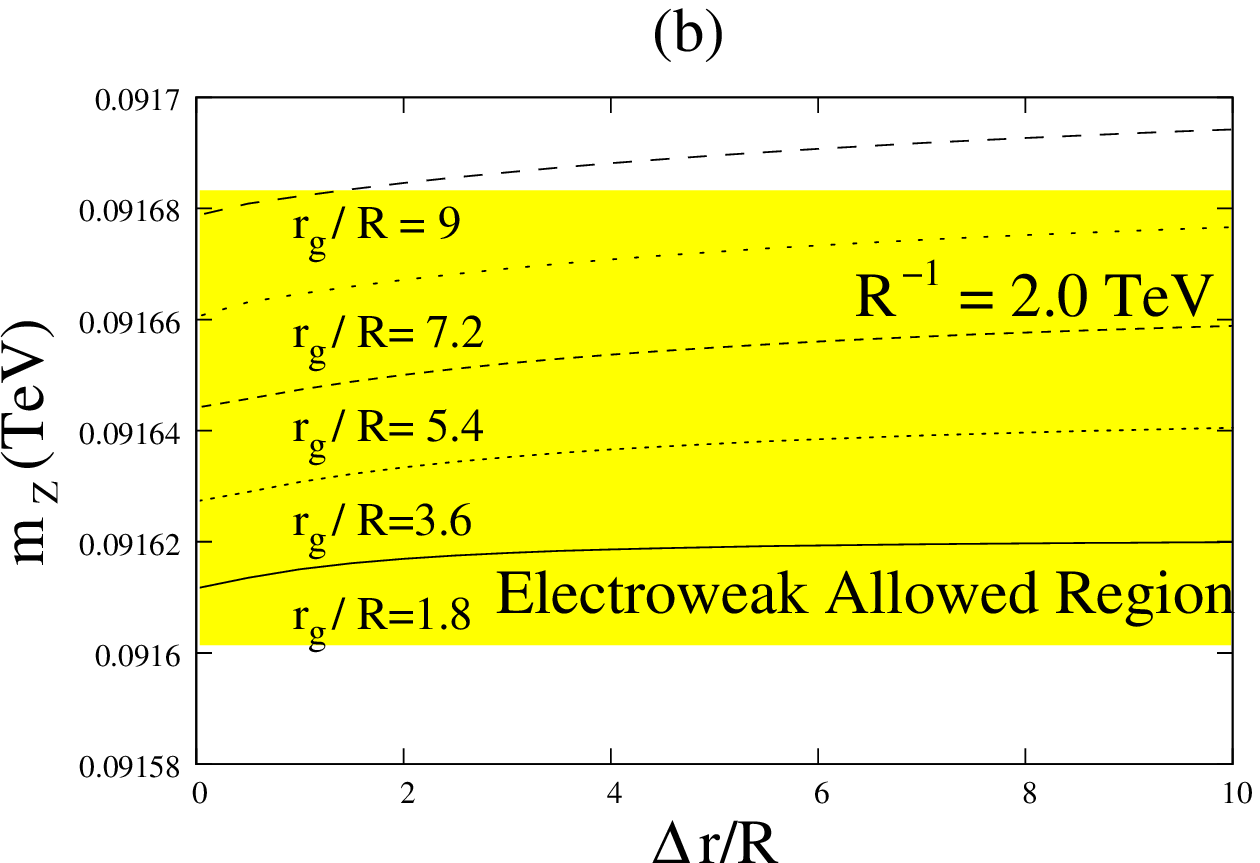}}
\end{center}
\caption{Constraints on $r_g/R$ and $\Delta r /R\equiv (r_H-r_g)/R$ from tree
level corrections to the $Z$ mass for (a) $R^{-1}=1$~TeV and (b)
$R^{-1}=2$~TeV.
We plot the lower edge of the $m_Z^{nUED}(R,r_g,r_H)$ band for different values
of $r_g/R$. The yellow (gray) band corresponds to the 2$\sigma$ allowed values
of $m_Z^{tree}$ as defined in Eq.~(\ref{mztreedef}). For a given value of
$r_g/R$ the bound on $\Delta r$ can be inferred from the intersection of the
corresponding contour line with the upper bound of the yellow (gray) band.}
\label{FigrHmz}
\end{figure}
Again using the electroweak values of  $\alpha = 1/(127.925 \pm 0.016)$,
$m_W = (80.398 \pm 0.025)$~GeV and $G_f = (11.66367 \pm 0.00005)$~TeV$^{-2}$
we can predict the tree-level $m_Z^{nUED}$ value.

Fig.~\ref{FigrHmz} shows the resulting predictions of the lower edge of the
$m_Z^{nUED}(R,r_g,r_H)$ band for compactification scales of $R^{-1}=1\mbox{ and } 2$ TeV.
The different contours correspond to fixed values of $r_g/R$ and varying
values of $\Delta r/R\equiv (r_H-r_g)/R$.
The yellow (gray) band corresponds to the 2$\sigma$ allowed values of
$m_Z^{tree}$. For a given value of $r_g/R$ the bound on $\Delta r$ can be
inferred from the intersection of the corresponding contour line with the
upper bound of the yellow (gray) band. As can be seen,
for moderate values of $r_g$ substantial splittings between $r_g$
and $r_H$ are allowed.

The analysis presented here can be extended to other Standard Model observables
which get modified at tree level. For example from Eqs.~(\ref{gw4Dto5Dtr}) and
Eq.~(\ref{gw4Dto5Dqu}) it can be seen that the $WWZ$ vertex and the $WWWW$
vertex are modified. We performed the full analysis for the $WWZ$ vertex and
found that it leads to constraints which are substantially weaker than the
constraints from $m_Z$ presented above. This is to be expected as the
experimental precision on $\alpha,G_f,m_W,m_Z$ is much higher than the
precision on the $WWZ$ vertex \cite{PDG}. Similarly, we expect the bounds from
the $WWWW$ vertex to be sub-dominant.

\subsubsection{Determining the LKP}\label{rHDM}

With $(\gy,\gw,\vh)$ fixed by matching to the Standard Model parameters for a
given set of $(r_g,r_H,R)$, the mass determining equations in
Eq.~(\ref{nQcondHBLKT}) fix the KK spectrum of all particles in the electroweak
sector. From  Eq.~(\ref{nQcondHBLKT}) it follows that
\bea
m_{Z^{(1)}}&\geq&m_{a^{0(1)}}>m_{a^{\pm(1)}}\nonumber\\
m_{Z^{(1)}}&>&m_{W^{\pm(1)}}\geq m_{a^{\pm(1)}}\nonumber\\
m_{W^{\pm(1)}}&>&m_{\gamma^{(1)}}
\eea
leaving $\gamma^{(1)},a^{\pm(1)}$ and $h^{(1)}$ as possible LKPs.
From Section~\ref{secrBeqrWth}, the mass of the KK photon
is determined by
\bea
r_g m_{\gamma^{(1)}} = \cot \frac{m_{\gamma^{(1)}} \pi R}{2}.
\eea
From Section~\ref{secChargedSpec}, the mass of the $a^{\pm(1)}$ is
\beq
m_{a^{\pm (1)}}^2 = \frac{\gw^2 \vh^2}{4} + (M_{a^{\pm(1)}})^2,
\eeq
where $M_{a^{\pm (1)}}$ satisfies the equation
\beq
r_H M_{a^{\pm(1)}} = \cot  \frac{M_{a^{\pm(1)}} \pi R}{2}.
\eeq

\begin{figure}
\begin{center}
\resizebox{15.cm}{!}{\includegraphics{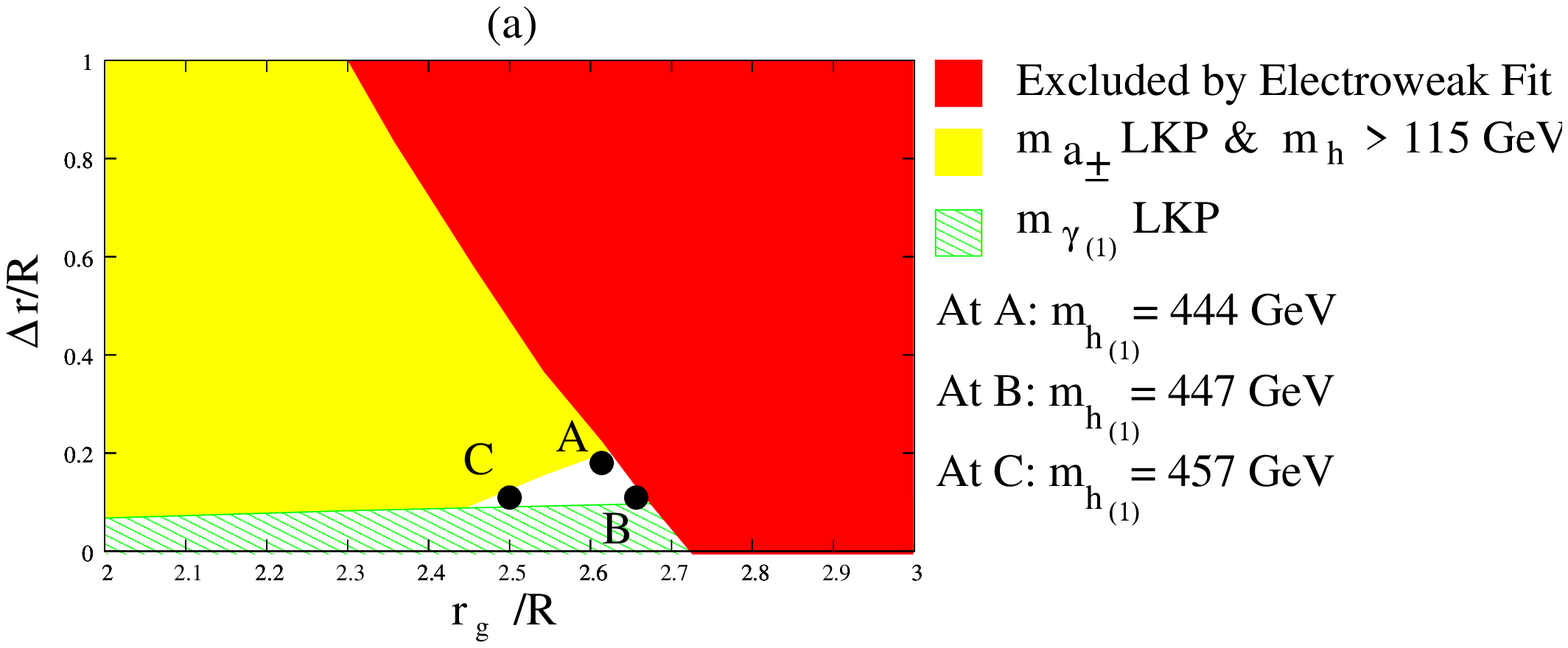}}
\resizebox{15.cm}{!}{\includegraphics{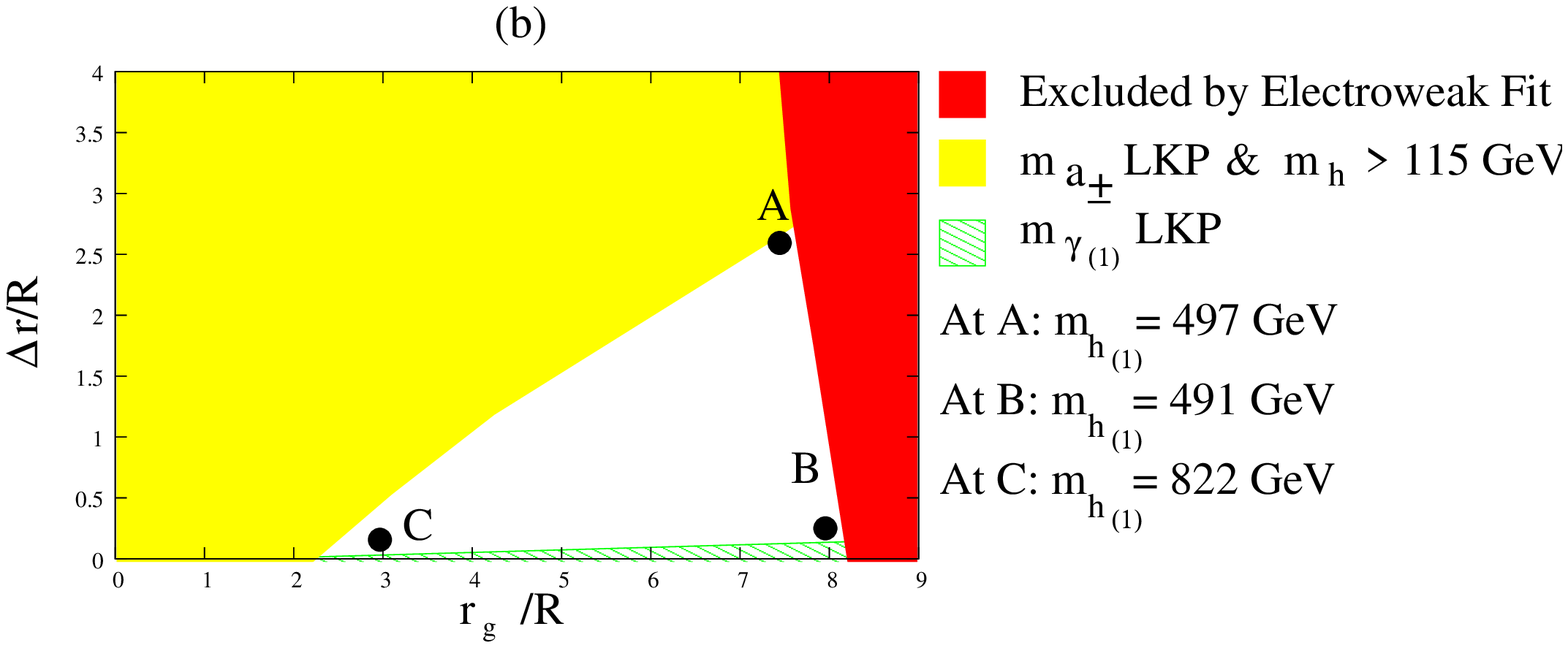}}
\end{center}
\caption{LKP phase space in the $r_g\equiv r_B=r_W\neq r_H$ scenario.
For (a) $R^{-1}=1$~TeV and (b)  $R^{-1}=2$~TeV, we plot the regions
of different types of LKP in the $\Delta r/R\equiv (r_H-r_g)/R$ versus
$r_g/R$ plane. The Higgs brane mass parameter has been set to $\mu_b=0$ and the zero mode
Higgs mass has been set to $115$~GeV. The red (dark gray) region is excluded
by the electroweak constraints in Section~\ref{SecrHLEPconstraints}. In the
yellow(gray) region the LKP is the $a_{\pm^{(1)}}$ while in the green (shaded)
region the KK photon is the LKP. In the white region the LKP is the KK Higgs.
The points $A$, $B$ and $C$ in each plot represent sample points with a Higgs
LKP whose mass is given on the right of the plots.}
\label{FigmhDM}
\end{figure}

While all other KK spectra in the electroweak sector are fixed, the KK spectrum
 of the Higgs depends on $\muh$ and $\mu_b$. In order to be able to compare the
 masses of the $a^{\pm(1)}$ and the $h^{(1)}$, we assume $\mu_b=0$ and fix
$\muh$ by demanding that the zero mode Higgs mass is fixed at $m_h=115$~GeV.
With these choices, the mass of the KK Higgs $m_{h^{(1)}}$ satisfies the
constraint
\beq
\frac{r_H m^2_{h^{(1)}}}{\sqrt{m_{h^{(1)}}^2 - 2 \muh^2}} = \cot
\left(\frac{\sqrt{m_{h^{(1)}}^2 - 2 \mu^2} \pi R}{2} \right)\label{homeq},
\eeq
where $\muh$ is determined from
\bea
\frac{r_H m_h^2}{\sqrt{2\muh^2-m_h^2}} &=& \tanh \left(\frac{\sqrt{
2\muh^2-m_h^2} \pi R}{2} \right).\label{hzmeq2}
\eea
Using these equations, we calculate the KK masses for the $h^{(1)}$,
$a^{\pm(1)}$ and $\gamma^{(1)}$ to find the LKP in the parameter space $(r_g,
\Delta r)$. Fig.~\ref{FigmhDM} shows the resulting LKP phase diagrams for
$R^{-1}=1\mbox{ and } 2$ ~TeV. The red (dark gray) areas are excluded by the
electroweak fit derived in the last section. For negative $\Delta r$ (not
displayed) and in the green (shaded) region at low $\Delta r$ the LKP is the KK
photon. In the yellow (gray) area either the LKP is the $a^{\pm(1)}$ or the
zero mode Higgs mass is less than $115$~GeV. Therefore the yellow (gray) region
is experimentally disfavored. The white triangular area
signifies the maximally allowed parameter space with a Higgs LKP. Assuming
that $\mu_b\neq 0$ or $m_h>115$~GeV makes the first KK Higgs heavier and
thus reduces the Higgs LKP parameter space further.

While for $R^{-1}=1$~TeV the Higgs LKP parameter space is strongly constrained,
 it opens up for a larger compactification scales as can be seen in
Fig.~\ref{FigmhDM}(b). We emphasize again that a higher compactification scale
does not imply a substantially heavier LKP. In Fig.~\ref{FigmhDM}(a)~and~(b), we give
 three sample points at the corners of the Higgs LKP parameter space with
their respective $h^{(1)}$ masses. The Higgs LKP mass for $R^{-1}=1$~TeV lies in a
range of $440\mbox{ GeV}\lesssim m_{h^{(1)}}\lesssim 460\mbox{ GeV}$, while
for $R^{-1}=2$~TeV the allowed Higgs LKP mass lies in the range of
$490\mbox{ GeV}
\lesssim m_{h^{(1)}}\lesssim 830\mbox{ GeV}$.

\begin{figure}
\begin{center}
\resizebox{7.5cm}{!}{\includegraphics{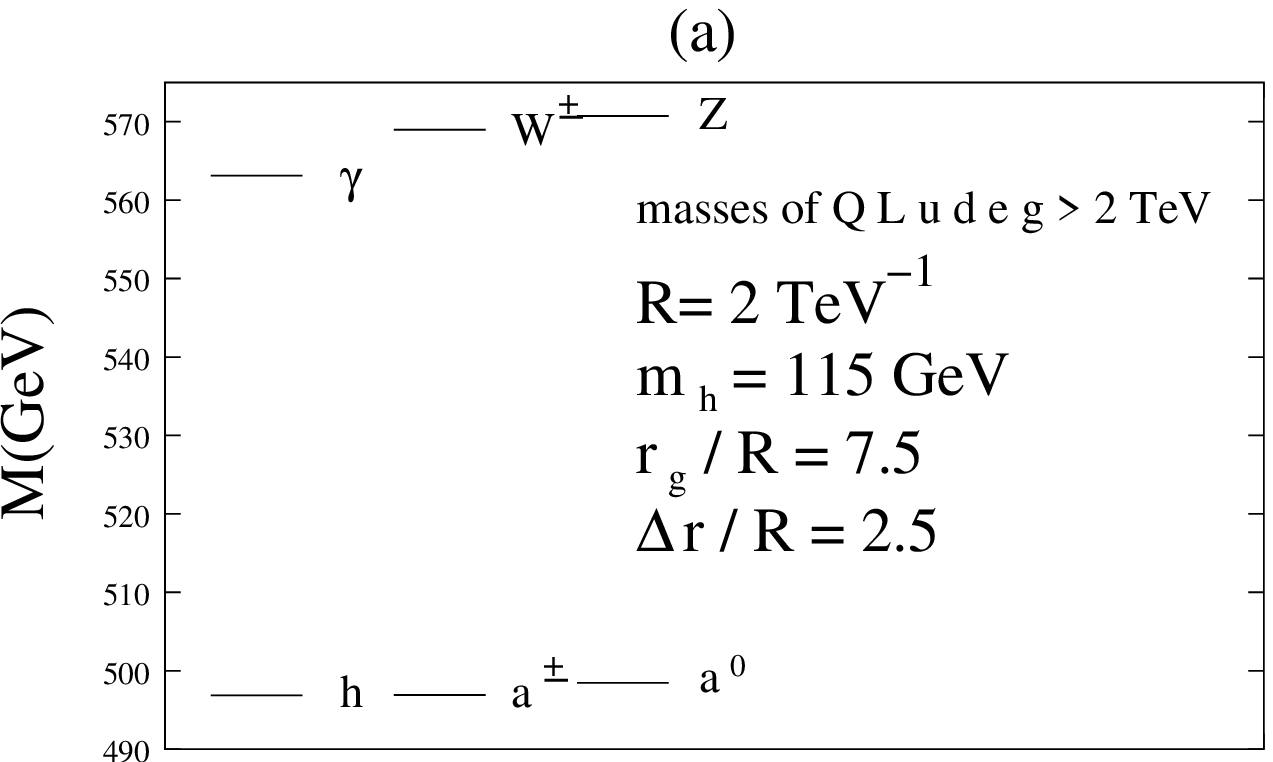}}
\resizebox{7.5cm}{!}{\includegraphics{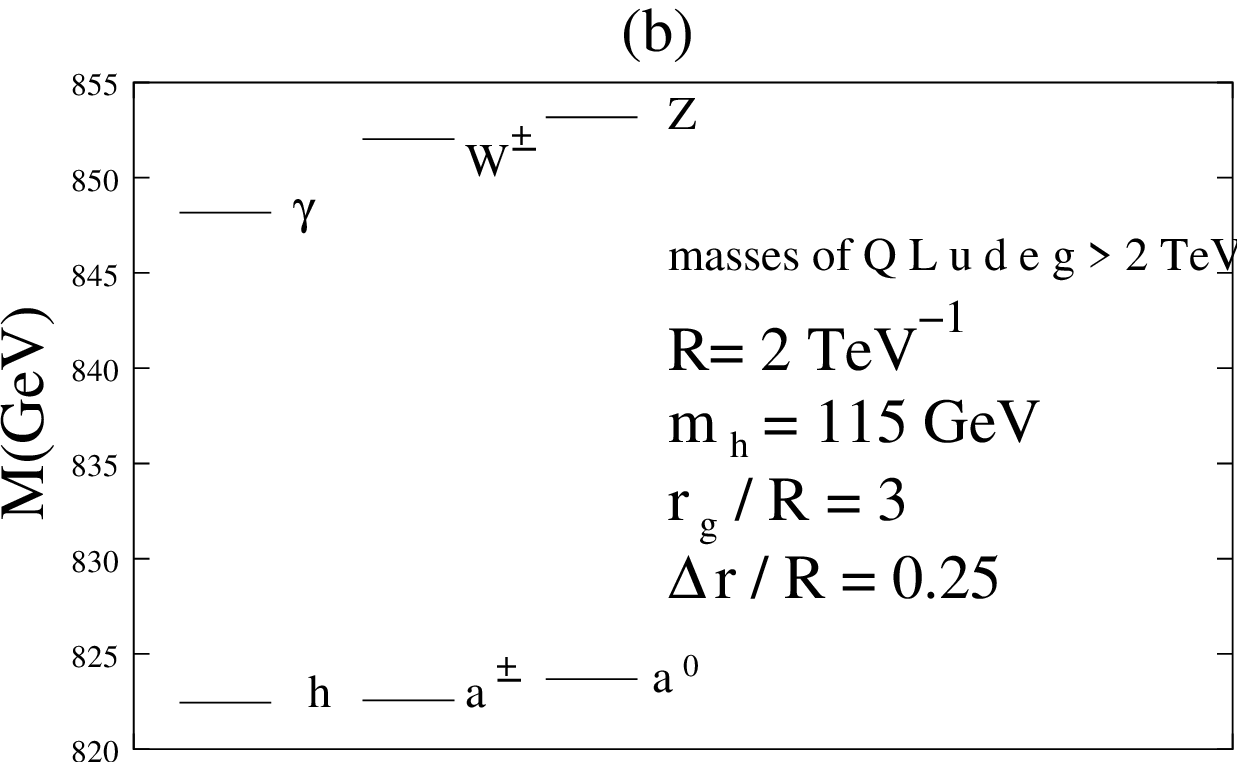}}
\end{center}
\caption{Sample spectra for UED with $r_g \neq r_H$.  (a) The tree level UED spectrum
with $R^{-1}=2$~TeV, $r_g \equiv r_B = r_W= 7.5 R$ and $\Delta r\equiv r_H-r_g= 2.5 R$ which corresponds to sample point A of Fig.~\ref{FigmhDM}(b). (b) The tree level UED spectrum
with $R^{-1}=2$~TeV, $r_g = 3 R$ and $\Delta r = 0.25 R$ which corresponds to sample point C of Fig.~\ref{FigmhDM}(b).}
\label{figmhspec}
\end{figure}

In Fig.\ref{figmhspec} (a) and (b), we show the first KK level masses of the sample points A and C for $R^{-1}=2$~TeV as defined in Fig.\ref{FigmhDM} (b). As is expected, the masses of $a^{\pm(1)}$ and $h^{(1)}$ at these points are almost degenerate. The masses of the KK gauge bosons lie higher because $r_H>r_g$. Comparing sample points A and C, point A lies at larger $r_H$ and $r_g$. Therefore the mass scale for all electroweak particles is lower than at point C. As $r_H-r_g$ is larger at point A, the relative splitting between the Higgs KK masses and the gauge boson KK masses is larger than at point C. The fermion and gluon KK mode masses remain at the compactification scale.  

\subsection{Scenario III: $r_w \neq r_B$}\label{secPhenoBneqW}

As a third example we investigate a scenario with $r_B\neq r_W$. As we showed,
boundary kinetic terms reduce the KK masses. Therefore setting $r_W>r_B$ is an
easy way to obtain a  $W^{3(1)}$-like LKP. To see qualitatively how the LKP is
changed in this scenario, consider the mass matrix at the first KK level
\beq
M^2_{B^{(1)},W^{3(1)}}=\left(
\begin{array}{cc}
m_{B^{(1)}}^2 & \mathcal{M}^2_{1,1}\\
 \mathcal{M}^2_{1,1} & m_{W^{(1)}}^2
\end{array}
\right),
\label{NGBMKKone}
\eeq
where the mixing elements $M^2_{m,n}$ have been defined in Eq.~(\ref{wbmixterm}). To
a good approximation, the LKP is the lighter eigenstate of this $B^{(1)}-
W^{3(1)}$ system. If we choose a sufficiently large $r_W$ then we can make
$ m_{W^{(1)}}^2< m_{B^{(1)}}^2$, in which case $W^{3(1)}_\mu$ becomes the main
component of lighter eigenstate of the  $B^{(1)}_\mu-W^{3(1)}_\mu$ system.

However this is not the full story. As indicated in Sec.~\ref{secBneqW} for $r_W\neq r_B$, the KK basis $\{f^{B}_n \}$ and $\{f^{W}_n \}$ differ and electroweak symmetry breaking induces $B-W^{3}$ mixing between different KK levels as given in Eq.~(\ref{wbmixterm}).

To address the KK mode mixing qualitatively, consider the only non-vanishing mass matrix elements which mix $B$ and $W^{3}$,\footnote{We give the argument for even KK modes, here. The same holds true for odd KK mode mixing. Mixing between even and odd modes is forbidden by KK parity.}
\beq
M^2_{B^{(2m)},W^{3(2n)}}=\left(
\begin{array}{cc}
m^2_{B^{(2m)}} & \mathcal{M}^2_{2m, 2n}\\
 \mathcal{M}^2_{2m, 2n}              & m^2_{W^{3 (2n)}}
\end{array}
\right).
\eeq
In the presence of BLTs, the mass difference is of the order
\beq
m^2_{B^{(2m)}}-m^2_{W^{3 (2n)}} \sim \left(\frac{2n}{R}\right)^2-\left(\frac{2m}{R}\right)^2=
\frac{4(m^2-n^2)}{R^2}
\eeq
if $m \neq n$, but the normalization of the wavefunctions in the extra dimension puts a bound on these mixing terms
\beq
\mathcal{M}^2_{2m, 2n}<\frac{\gy\gw\vh^2}{4}.
\eeq
Hence the KK mode mixing between the $B^{(n)}$ and  $W^{(m)}$ modes satisfies
\beq
\sin^2 \theta_{mn}\lsim \frac{\left(\frac{\gy\gw\vh^2}{4}
\right)^2}{\frac{4(m^2-n^2)m^2}{R^4}},
\eeq
which is small whenever $R^{-1}>>\frac{\gy\gw\vh^2}{4}$ and $m \neq n$. The largest mixing of modes between the
$B$ and $W$ KK towers occurs when $m = n$, and to good approximation, the mass matrix can be diagonalized KK level by KK level, underlining the validity of the qualitative argument on a $W^{3(1)}$ LKP given above. We perform a quantitative study of obtaining a $W^{3(1)}$ LKP via BLTs in Section~\ref{secRwDM} in which we take the KK mode mixing into account numerically.

From the above discussion it is clear that the zero modes mix with higher KK modes as well. Matching the zero mode sector on to the
Standard Model is therefore non-trivial. In the next section, we numerically diagonalize the neutral gauge sector mass matrix and follow the same procedure as in Sections~\ref{rewcon} and \ref{SecrHLEPconstraints} in order to determine electroweak constraints on $r_W$.

\subsubsection{Electroweak constraints}\label{secChargeVio}

To perform a quantitative study of the LKP we choose the special case
$r_W\neq 0=r_B=r_H$.  As outlined in Section~\ref{secBneqW} the equations
of motion and boundary conditions for $W$ are
\bea
\left[\left(\Box-\p_5^2+\frac{\gw^2\vh^2}{4}\right)\eta^{\mu\nu}-\p^\mu \p^\nu
\right]W^3_\nu &=& 0 \label{W3EOM}\\
\left(\pm\p_5+r_W\left[\Box \eta^{\mu\nu}-\p^\mu\p^\nu\right]\right) W^3_\nu |_{y=\pi R,0} &=&0\label{W3bdvar}
\eea
and for $B$ are
\bea
\left[\left(\Box-\p_5^2+\frac{\gy^2\vh^2}{4}\right)\eta^{\mu\nu}-\p^\mu \p^\nu
\right]B_\nu &=& 0 \label{BEOM}\\
\pm\p_5 B_\nu |_{y=\pi R,0} &=&0.\label{Bbdvar}
\eea
Hence we expand $W^{(3)}_\mu$ in the basis for a gauge field with
BLKTs and (as we chose $r_B=0$) we expand $B_\mu$ in the ``standard'' UED
basis. However, as $W^{(3)}_\mu$ and $B_\mu$ are not the mass eigenstates of
the bulk Lagrangian with electroweak symmetry breaking,
the term $\vh^2 W^\mu B_\mu$ induces the mass mixings given in
Eq.(\ref{wbmixterm}).

For the charged gauge field sector, from Eq.~(\ref{nQcondHBLKT}) the zero mode
mass is $m_W^{(0)}$ is determined by
\beq
\frac{r_W m^2_{W^{(0)}}}{\sqrt{m^2_{W^{(0)}}+\mwh^2}} = \tanh \frac{
\sqrt{m^2_{W^{(0)}})+\mwh^2}\pi R}{2} \label{WZquantrW}.
\eeq
Turning this relation around we can determine $\mwh$ by setting $m_W^{(0)}\equiv m_W$.
With $\mwh$ determined, the mass spectrum of $W^{\pm(n)}$ is fixed. As in Sections~\ref{rewcon}
and \ref{SecrHLEPconstraints}, $\gw$ can be determined by calculating the effective Fermi constant
from Eq.~(\ref{GfdefrH}), where the non-zero $W^\pm$ KK mode contributions $b_{2n}$ are given by Eq.~(\ref{bcoeffdef}) with $r_{EW}\to r_W$ 
while the hyperbolic zero mode implies 
\beq
b_{0} = \frac{8\sinh^2 \frac{M_{2n}^W \pi R}{2}}{
\left(1 + \frac{\sinh M_{2n}^W \pi R}{M_{2n}^W \pi R} +\frac{4 r_W}{\pi R}
 \cosh^2 \frac{M_{2n}^W \pi R}{2}\right) \left(M_{2n}^W \pi
R \right)^2}.
\eeq

The above results enable us to determine $\gw$ and $\vh$ from the charged sector for a given set of parameters $R,r_W$.
The next goal is to calculate $\alpha$ and use it as a matching condition for $\gy$. In Sections~\ref{rewcon}
and \ref{SecrHLEPconstraints}, we were able to give an explicit and easily invertible relation for $\alpha(R,r_g,r_H,\gy,\gw)$.
For $r_W\neq r_B$, this task is complicated by the fact that the photon is only determined implicitly as the lightest eigenstate of the neutral gauge
boson mass matrix.

The relationship between the 4 dimensional mass
eigenstates and the original $B - W$ basis is
\bea
A^{(0)}&=&U^T_{1j}\mathcal{B}^j \label{rWphotonID}\\
Z^{(0)}&=&U^T_{2j}\mathcal{B}^j\label{rWZID}\\
A^{(1)}&=&U^T_{3j}\mathcal{B}^j,\label{rWLKPID}\\
Z^{(1)}&=&U^T_{4j}\mathcal{B}^j,\label{rWZKKID}\\
&\vdots& \nonumber
\eea
where $\mathcal{B}^j$ denotes $(B^{(0)},W^{3(0)},B^{(1)},W^{3(1)}, ...)^T$.

In order to determine the basis transformation $U_{ij}$. We start from a trial value of $\gy$.  Together with the already determined values of $\gw$ and $\vh$, we calculate the neutral gauge boson mass mixing matrix.
As we work in the $B-W$ basis, the $W^{3(m)}-W^{3(n)}$ and $B^{(m)}-B^{(n)}$ mixing terms are zero for $m\neq n$.  The diagonal elements are given by $m^2_{W^{(n)}}$ and  $m^2_{B^{(n)}}$, while the $B^{(n)}-W^{(m)}$ mixing is described in Eq.~(\ref{wbmixterm}). We numerically diagonalize the trial mass matrix to obtain $U_{ij}$. Eq.~(\ref{rWphotonID}) then determines the photon in the $B-W$ basis. The couplings of the zero mode
photon to zero mode fermions can be derived from the 5 dimensional action
\bea
S&\supset&\int d^4x dy \,i \,\bar{e}_R\gamma^M B_M e_R\supset \int d^4x \,i \,Q_{e_R}\bar{e}_R^{(0)}\gamma^\mu A_\mu e_R^{(0)}\label{eRcoupling}\nonumber\\
S&\supset& \int d^4x dy\, \frac{i}{2} \bar{e}_L\gamma^M B_M e_L +\frac{i}{2}\bar{e}_L\gamma^M {W}^3_M e_L \label{eLcoupling}\nonumber\\
 &\supset&\int d^4x\, i \,Q_{e_L} \bar{e}_L^{(0)}\gamma^\mu A_\mu^{(0)}e_L^{(0)},\nonumber\\
\eea
to be
\bea
Q_{e_R}&=&\sum_{j=0}^{\infty}\int dy\, f^{e_R}_0 U^T_{1(2j+1)}f^B_{j}f^{e_R}_0
\label{QRdef}\\
Q_{e_L}&=&\frac{1}{2}\sum_{j=0}^{\infty} \left(\int dy\, f^{e_L}_0 U^T_{1(
2j+1)} f^B_j f^{e_L}_0+ \int dy f^{e_L}_0 U^T_{1(2j+2)}
f^W_j f^{e_L}_0 \right).\label{QLdef}
\eea
With the electric charge of the fermions determined, we calculate $\alpha=Q^2/4\pi$. We iterate this procedure with varying values of $\gy$ until the result matches the experimentally measured value $\alpha_{em} = 1/(127.925 \pm 0.016)$ to a precision of $(\alpha-\alpha_{em})/\alpha_{em}<10^{-7}$ which lies an order of magnitude below the experimental error.\footnote{The equality of left-handed and right-handed couplings is guaranteed by 4D gauge invariance of the \emph{full} model. In order to perform the diagonalization numerically, we truncate the mass matrix after the fourth KK level. The truncation leads to numerical values of $(\alpha_L-\alpha_R)/\alpha_L < 10^{-7}$ for the parameter regime of $R,r_W$ considered in this article which is sufficiently low to not affect any of our results and justifies our truncation at the fourth KK level. As a consistency check, we verified numerically that $(\alpha-\alpha_{em})/\alpha_{em}$ decreases when truncating at a higher KK level. In our determination for $\gy$, we use $Q_{e_L}$ to calculate $\alpha$.}
With the parameters $(\gy,\gw,\vh)$ determined, we calculate the mass of the second lightest mass eigenstate which is identified with the tree level $Z$ mass as defined in Eq.~(\ref{mztreedef}).

\begin{figure}
\begin{center}
\resizebox{7.5cm}{!}{\includegraphics{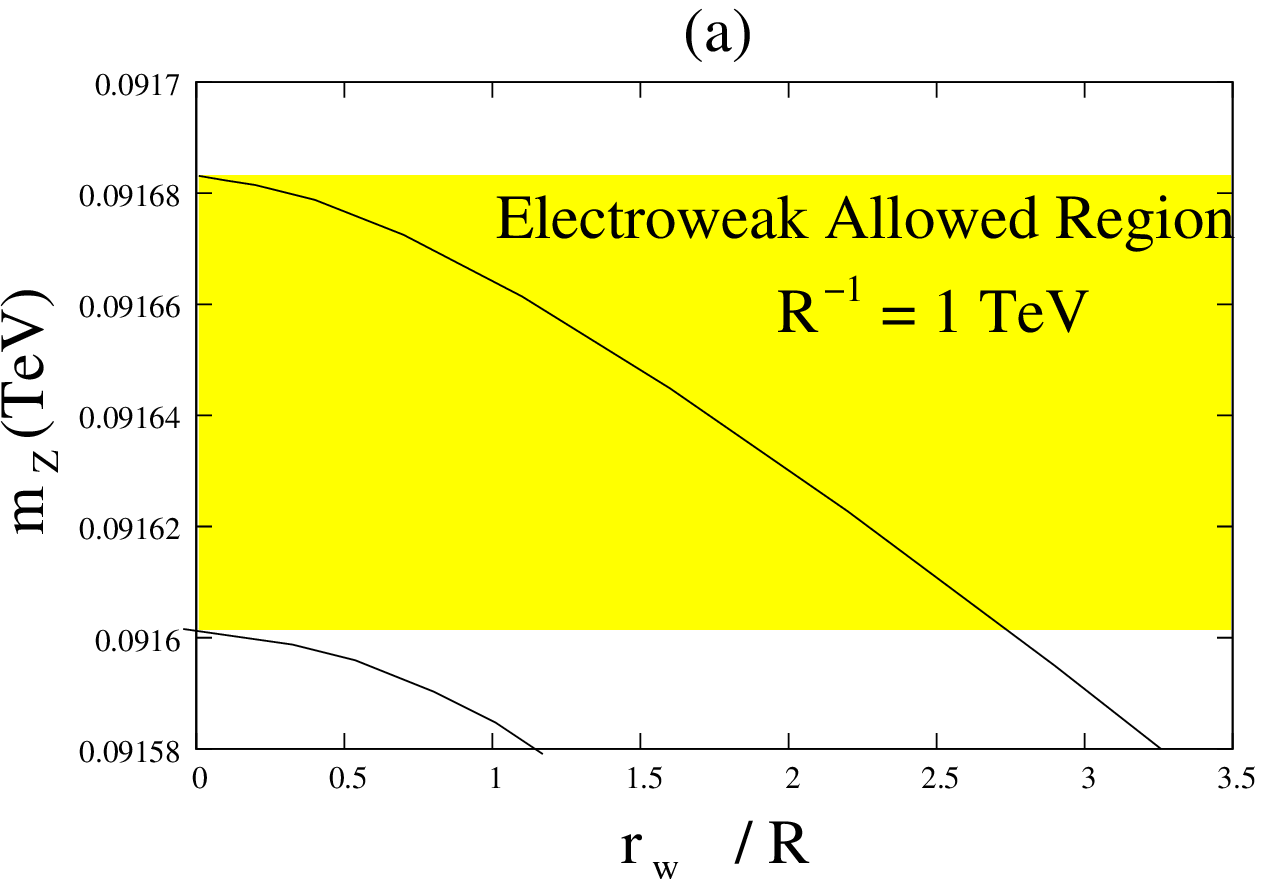}}
\resizebox{7.5cm}{!}{\includegraphics{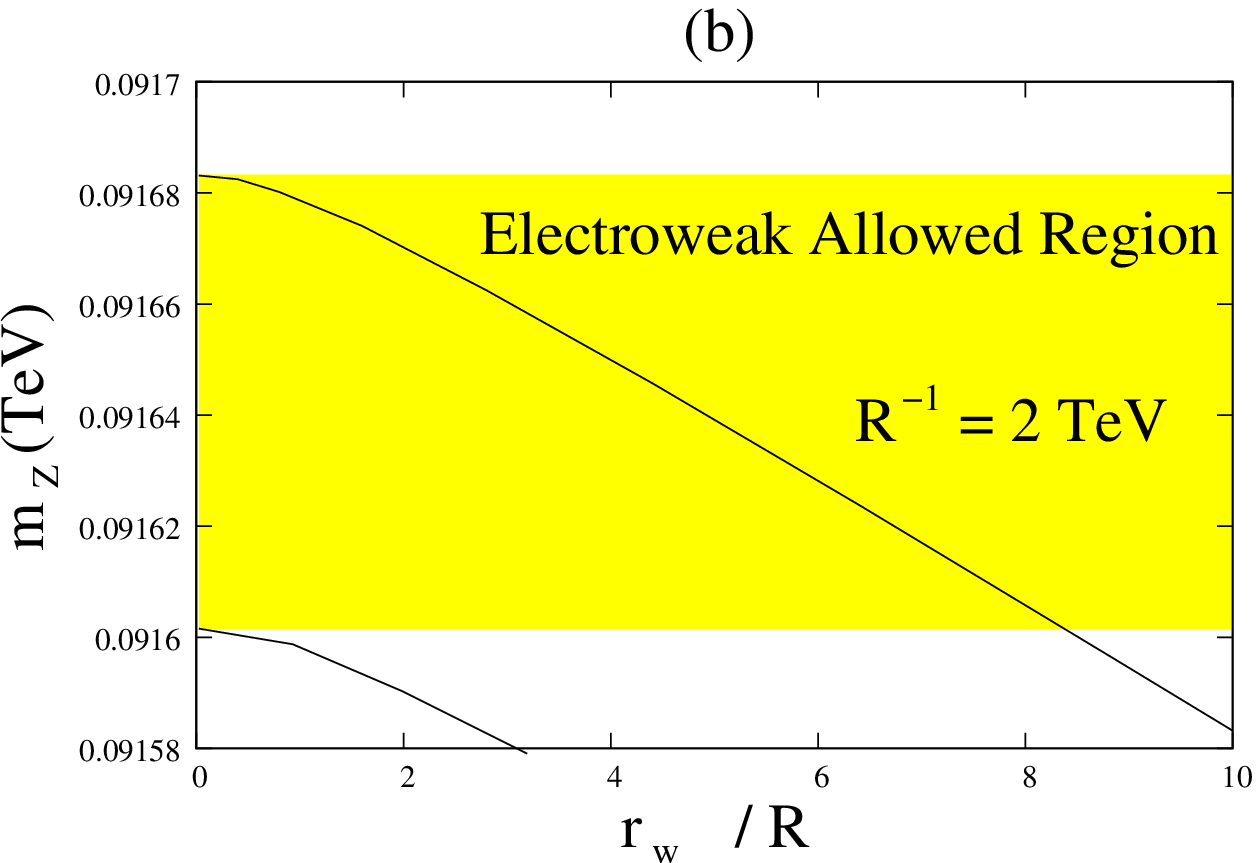}}
\end{center}
\caption{Variation of $m_Z^{nUED}$ for different values of $r_W$
with $R^{-1}=1$~TeV and $R^{-1}=2$~TeV in the $r_W\neq 0=r_B=r_H$ scenario. The yellow (gray) band
corresponds to the 2$\sigma$ allowed tree-level value of $m_Z^{tree}$.
The region within the black lines is the 2$\sigma$ predicted tree-level value
of $m_Z^{nUED}$.}
\label{Figrw}
\end{figure}

In Fig.~\ref{Figrw} we show the value of $m_Z^{nUED}$ for $R^{-1}=1$~TeV
and $R^{-1} = 2$~TeV, assuming a 2$\sigma$ error in the input values of
$\alpha$, $m_W$ and $G_f$.
As in scenarios I and II we see that the bounds on $r_W/R$ are weaker for increasing $R^{-1}$, and we again find that the bounds from tree level matching of the lowest lying modes to the Standard Model fields are weaker than the NDA estimate of $r_W/R\lsim 6 \pi/ \Lambda R$. As for scenario II, we performed the analysis for the $WWZ$ vertex modifications and we again find that it leads to weaker constraints than those from $m_Z$ presented above.

\subsubsection{$W^{3(1)}$ dark matter}\label{secRwDM}

In the last section, we determined the 5 dimensional parameters $(\gy,\gw,\vh)$ for a given set of parameters $R,r_W$ and calculated the bound on $r_W/R$ from matching the lowest lying KK modes to the standard model. With the parameter set $(r_W,R,\gy,\gw,\vh)$, the full KK mass spectrum of the electroweak sector is determined and we can study the LKP.\footnote{Again, in order to fully fix the Higgs boson KK spectrum, $m_h$ and $\mu_b$ need to be specified, but assuming $m_h>115$~GeV or $\mu_b>0$ raises the Higgs KK masses without affecting the mass spectra of the electroweak gauge bosons or the charged and the pseudoscalar Higgs. The discussion of the LKP is therefore independent of $m_h$ and $\mu_b$.}

For $r_W\neq 0 = r_B = r_H$ it follows from Eq.~(\ref{nQcondHBLKT})
\bea
m_{h^{(1)}}&>&m_{a^{0(1)}}>m_{a^{\pm(1)}}>m_{W^{\pm(1)}}\nonumber\\
m_{Z^{(1)}}&>&m_{W^{\pm(1)}}\geq m_{A^{(1)}},
\eea
where $A^{(1)}$ and $Z^{(1)}$ denote the lightest and second lightest KK parity odd eigenvalue of the neutral gauge boson mass mixing matrix. In terms of the $B-W$ basis, they are defined in Eqs.~(\ref{rWLKPID}) and (\ref{rWZKKID}) as $A^{(1)}=U^T_{3j}\mathcal{B}^j$ and  $Z^{(1)}=U^T_{4j}\mathcal{B}^j$. As argued in the beginning of Section~\ref{secPhenoBneqW}, the KK mode mixing contributions in this basis are suppressed and hence all  $A^{(1)}$ components except $U^T_{33}B^{(1)}$ and $U^T_{34}W^{(1)}$ are sub-dominant. It is thus sensible to define the Weinberg angle at the first KK level by
\beq
\tan \theta_W^{(1)}\equiv U^T_{34}/U^T_{33}.
\label{defKKWeinberg}
\eeq
\begin{figure}
\begin{center}
\resizebox{9.cm}{!}{\includegraphics{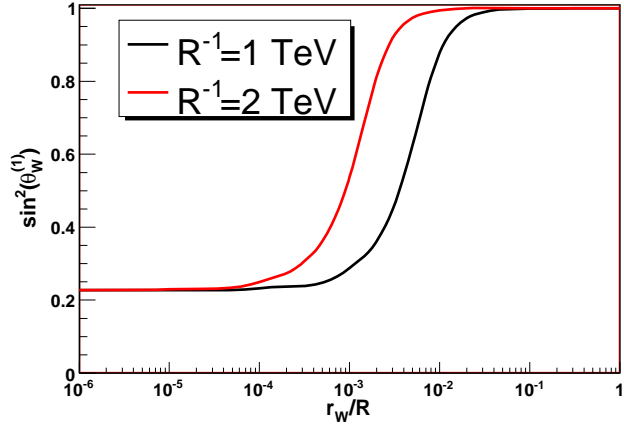}}
\end{center} 
\caption{Modification of the Weinberg angle of the first KK mode as defined in Eq.~(\ref{defKKWeinberg}) at $R^{(-1)}= 1$ TeV (black) and $R^{(-1)}= 2$ TeV (red/gray).}
 \label{FigrWneqrB}
\end{figure}

Fig.~\ref{FigrWneqrB} shows the numerical result for $\sin^2(\theta^{(1)}_W)$ as a function of $r_W$, calculated for $R^{-1}=1\mbox{ and }2$~TeV. As can be seen, for values of $r_W/R\sim\mathcal{O}(10^{-2})$ ($r_W/R\sim\mathcal{O}(10^{-3})$) for $R^{-1}=1$~TeV ($R^{-1}=2$~TeV), the LKP composition changes from a KK photon with Weinberg angle $\theta^{(1)}_W= \theta_{W,SM}$ to almost purely $W^{3(1)}$. The bounds on $r_W/R$ derived in the last section lie two or three orders of magnitude above these values, showing that $W^{3(1)}$ dark matter can easily be accomplished in the non-minimal UED models studied in this article.

As in scenario I and II, the first KK mode masses in this scenario are split. The fermion and gluon KK mode masses lie at the compactification scale $R^{-1}$. As we chose $r_H=0$, the masses of the Higgs KK modes $h^{(1)}, a^{\pm,0(1)}$ lie at the compactification scale, too. Furthermore, the choice $r_B=0$ leaves the mass of the heavier first KK level eigenstate of the neutral gauge boson mass matrix at the compactification scale. For $r_W/R\gtrsim 10^{-2}$, the LKP is almost purely $W^{3(1)}$. As is expected, only the masses of the $W^{\pm,3 (1)}$ are reduced by the non-zero $SU(2)$ BLKT. As the mass splitting between $W^{\pm (1)}$ and $W^{3(1)}$ solely arises from the mixing in the neutral gauge boson sector, which is strongly suppressed for $r_W/R\gtrsim 10^{-2}$, the $W^{\pm (1)}$ and $W^{3(1)}$ masses are highly degenerate in our tree-level investigation.

We wish to emphasize that the analysis presented here is performed at tree-level, only. In the discussion of radiative corrections to MUED in Ref.~\cite{Cheng:2002iz}, BLTs are assumed to vanish at the cutoff scale $\Lambda$. This 
assumption is used to justify neglecting the effect of KK-mode mixing on the 
mass spectrum and the modification of the KK-eigenfunctions. In this article 
we have shown that such modifications of the KK-eigenfunctions are a key 
aspect when considering non-vanishing BLTs. Therefore, a generalization of the results of Ref.~\cite{Cheng:2002iz} to the case of non-minimal UED is non-trivial. To give an estimate of the impact of radiative corrections, following  Ref.~\cite{Cheng:2002iz}, let us again consider the mass matrix of neutral gauge bosons at the first KK level in Eq.~(\ref{NGBMKKone}). Including radiative corrections, BLTs, and neglecting KK mode mixing, we can naively approximate the mass 
matrix to be
\beq
M^2_{B^{(1)},W^{3(1)}}=\left(
\begin{array}{cc}
m_{B^{(1)}}^2+\hat{\delta}\left(m^2_{B^{(1)}}\right) & \mathcal{M}^2_{1,1}\\
 \mathcal{M}^2_{1,1} & m_{W^{(1)}}^2+\hat{\delta}\left(m^2_{W^{3(1)}}\right)
\end{array}
\right),
\eeq
where the matrix elements are defined as in Eq.~(\ref{NGBMKKone}) and $\hat{\delta}(m^2_{B^{(1)}})$ and $\hat{\delta}(m^2_{W^{3(1)}})$ are the radiative corrections to $m^2_{B^{(1)}}$ and $m^2_{W^{3(1)}}$ as given in Ref.~\cite{Cheng:2002iz}. In MUED, the radiative corrections drive the LKP to be almost purely $B^{(1)}$ because $m_{B^{(1)}}^2=1/R^2+m_B^2$, $m_{W^{3(1)}}^2=1/R^2+m_W^2$ and  $\hat{\delta}(m^2_{B^{(1)}})<0<\hat{\delta}(m^2_{W^{3(1)}})$. To arrive at a $W^{3(1)}$-like LKP, the BLT effects have to (over-)compensate the effect of radiative corrections. For $R^{-1}=1$~TeV and $\Lambda R = 20$, the radiative corrections in Ref.~\cite{Cheng:2002iz} yield a splitting of $\sim 60$~GeV between the MUED $B^{(1)}$ and $W^{3(1)}$ masses which can compensated by a $W$ BLKT with  $r_W/R\gsim 0.1$.  This value lies an order of magnitude below the NDA estimate $r_W/R\lesssim 6\pi/(\Lambda R)\approx 1$ and below the experimental constraints derived in this article. Therefore our simple approximation shows that a $W^{3(1)}$-like dark matter candidate is still conceivable, when taking radiative corrections into account.

\section{Conclusions and Outlook}\label{secConclusion}

Models with a universal extra dimension should be considered as effective field theories with a cutoff at the PeV scale. Their symmetries allow for brane localized operators of the same dimension as the standard UED bulk operators. Every brane localized term implies an unsuppressed free parameter. We argue that in a bottom-up approach to UED, all boundary localized operators should be included and their implications for phenomenology studied.

In this article, we extended earlier results on boundary localized operators \cite{Dvali:2001gm,Carena:2002me} in extra dimensional models and applied them to UED. We presented a framework to derive mass spectra and couplings for tree level UED with boundary localized kinetic and mass terms in the electroweak sector, including gauge fixing and identifying the Goldstone and physical Higgs KK modes. As we showed, the zero mode relations are generically modified when including BLTs. As the zero mode level of UED is identified with the Standard Model particles, modifications of the zero modes imply corrections to the Standard Model relations between masses and couplings \emph{at the tree level}. In addition, BLTs affect the mass spectrum \emph{and} couplings of the non-zero KK modes and thus have important implications for the collider phenomenology of UED KK modes as well as for the phenomenology of the UED dark matter candidate which is given by the lightest Kaluza Klein mode.

In order to demonstrate the phenomenological impact of boundary localized parameters in more detail, we presented three sample scenarios in which we derived constraints on the boundary parameters $r_B,r_W,r_H$ which arise when matching the zero modes of the electroweak sector to the Standard Model. With the underlying parameters fixed by the matching we identified the lightest KK mode in the respective scenarios. In all sample scenarios studied, we showed that the inclusion of non-zero boundary kinetic terms change the 4D to 5D parameter matching, but once the correct relations are taken into account, the Standard Model relations are only weakly affected and lead to bounds on $r_W,r_B$ and $r_H$ which are substantially weaker than the NDA estimate of $r_B\sim r_W\sim r_H\sim 6 \pi/ (\Lambda R)$. On the contrary, non-zero boundary kinetic terms have a strong effect on the non-zero  KK masses and couplings of the electroweak sector. We showed that masses of the electroweak KK partners equipped with a boundary localized kinetic term can lie substantially below the compactification scale while the first KK modes of the fermions and the gluon remain at a mass scale $R^{-1}$.\footnote{With the content of this article, including a BLKT for the gluon is straight forward. This would reduce the KK gluon mass to the scales of the electroweak KK partners while still keeping the fermions heavy. This would realize a UED scenario which shares many qualitative features with split supersymmetry~\cite{splitsusy}.}

Our first sample scenario, where we chose $r_W=r_B=r_H$, represents a very simple example of non-minimal UED with only four free parameters $(R,r_{EW},m_h,\mu_b)$. In this scenario, the LKP is the KK photon for all allowed values of $r_{EW}$. For our second parameter choice, $r_B=r_W\equiv r_g\neq r_H$, we showed that the parameter space allowed by the electroweak constraints includes regions where the LKP is changed from a KK photon to either a KK charged Higgs $a^{\pm(1)}$ (which as a charged LKP is disfavored) or the KK Higgs $h^{(1)}$. The  KK Higgs provides a new dark matter candidate in UED models. For a compactification scale of $R^{-1}=1$~TeV, the $h^{(1)}$ LKP parameter space is strongly restricted, but it is opened up considerably for a compactification scale of $R^{-1}=2$~TeV. In our final scenario with $r_W\neq r_B=0=r_H$, we showed that a $W^{3(1)}$-like LKP can easily be achieved within the bounds on the zero mode spectrum studied in this article. $W^{3(1)}$-like dark matter presents another new UED dark matter candidate which is rarely considered in the literature, so far.\footnote{To our knowledge, the only more detailed investigation of $W^{3(1)}$-like dark matter and Higgs UED dark matter can be found in Ref.~\cite{Arrenberg:2008wy}, which assumes couplings of the LKP that are not modified by non-trivial wavefunction overlap integrals.} With the tools provided in this article (specifically in Appendix~\ref{appSpecs}), and the sample studies of Section \ref{secPheno}, detailed studies of models with other non-minimal parameters $(r_W,r_B,r_H,\mu_b)$ are straight forward. Beyond providing calculational examples, Section~\ref{secPheno} provides an instructive overview of the effects of BLTs on the electroweak sector of non-minimal UED at tree-level.

The work presented here is meant to be a small step towards a more complete mapping of the UED parameter space. We made the simplifying assumption that the boundary and bulk VEVs coincide which allowed us to expand around a flat VEV. For a general treatment, this assumption needs to be relaxed. We did not address boundary localized kinetic fermion terms in this article. All analysis presented in this article is at tree-level with BLTs only. On the contrary, the analysis of Ref.\cite{Cheng:2002iz} is at one-loop level but ignoring non-zero BLTs at the cutoff scale. For a complete treatment of UED as an effective field theory a consistent description of UED with loop corrections in the presence of BLTs is needed in order to study constraints on the \emph{full} UED parameter space along the lines of Refs.~\cite{UEDcolliderbounds,UEDprecisionbounds,UEDflavorbounds,UEDprecisionOther}. In the light of upcoming LHC data and dark matter searches, a more complete mapping of the UED parameter space is needed in order to investigate the potential of UED to explain new signals and also to allow for an honest comparison of UED with other Standard Model extensions, \eg along the lines of Refs.\cite{UEDvsSUSY}.

\vskip 0.5in
\begin{center}
{\bf Acknowledgments}
\end{center}
\vskip0.05in
We thank Marcela Carena, Gordon Kane, Kyoungchul Kong, Mariano Quiros, Carlos Wagner, James Wells, and especially Aaron Pierce for helpful discussions. This work was supported by the MCTP and the DOE under grant DE-FG02-95ER40899.

\vskip0.05in
\begin{center}
{\Large \bf Appendices}
\end{center}
\appendix

\section{Review and reformulation of electroweak symmetry breaking by a bulk Higgs on $S^1/\mathbb{Z}_2$}\label{appgaugefix}

In this appendix we reformulate the gauge fixing procedure of Ref.~\cite{Muck:2001yv} along the lines of Ref.~\cite{Cacciapaglia:2005pa}. This reformulation will enable us to generalize the gauge fixing procedure and the identification of the Goldstone modes and the physical Higgs modes to the electroweak sector of UED in the presence of boundary kinetic terms for the gauge and Higgs fields, which is presented in Appendix~\ref{appgaugefixbd}.

\subsection{The 5 dimensional abelian Higgs model}\label{subsecabelian}
In order to present the method in the simplest setup first, consider the UED action of an abelian gauge field on $S^1/\mathbb{Z}_2$, spontaneously broken by a bulk Higgs field
\beq
S_{EW,bulk}=S_g+S_H
\label{bulkactionfull}
\eeq
with
\bea
S_g &=& \int d^5x \left( -\frac{1}{4\g^2}A_{MN}A^{MN} \right) \\
S_H &=& \int d^5x \left( (D_MH)^{\dagger}(D^MH)+\muh^2H^{\dagger}H-\lambdah(H^{\dagger}H)^2 \right),
\label{bulkaction}
\eea
where the covariant derivative is $D_M=\partial_M-\frac{i}{2} A_M$. The 5 dimensional
Higgs boson acquires an vacuum expectation value (VEV) $\vh\equiv \sqrt{\muh^2/
\lambdah}$, and makes the 5 dimensional vector particle $A_M$ massive. Now we can expand
the 5 dimensional Higgs about the true vacuum
\beq
H = \frac{1}{\sqrt{2}} \left(v + h + i \chi \right)
\eeq
where $h$ is the physical scalar Higgs boson while $\chi$ is the pseudo-scalar
Higgs component. To eliminate
the bulk mixing term between the vector mode $A_\mu$ and the scalar modes $A_5$
and $\chi$ we can use the gauge fixing action
\bea
S_{GF}&=&\int \left( -\frac{1}{2\g^2\xi_A}\left[\p^\mu A_\mu-\xi_A\left(\p_5A_5-\frac{\g^2\vh}{2}\chi\right)\right]^2 \right),
\label{bulkgaugefix}
\eea
which helps us identify the Goldstone mode as
\bea
G &\equiv& \p_5 A_5 - \frac{\g^2\vh}{2}\chi. \label{Goldstone:eq}
\eea

Variation of the action yields the following equations of motion
\bea
& & \left[\Box-\p_5^2+2\muh^2\right]h = 0\label{hEOMab2}\\
& & \left[\left(\Box-\p_5^2+\frac{\g^2\vh^2}{4}\right)\eta^{\mu\nu}-\left(1-
\frac{1}{\xi_A}\right)\p^\mu \p^\nu \right]A_\mu = 0 \label{BEOMab2}\\
& & -\frac{1}{\g^2}\left(\Box-\xi_A \p_5^2+ \frac{\g^2\vh^2}{4}\right)A_5+
\frac{\vh}{2}(1-\xi_A)\p_5\chi = 0\label{B5EOMab2}\\
& & -(\Box-\p_5^2+\xi_A \frac{\g^2\vh^2}{4})\chi-\frac{\vh}{2}(1-\xi_A)\p_5
A_5 = 0\label{chiEOMab2}.
\eea
Subtracting $\g^2$ times Eq.~(\ref{B5EOMab2}) from $\p_5$ times
Eq.~(\ref{chiEOMab2}) eliminates the $\xi_A$ dependent part and allows us to
identify the physical pseudo-scalar~\cite{Muck:2001yv}
\bea
a &\equiv& -\p_5 \chi +\frac{\vh}{2} A_5, \label{physHiggs:eq}
\eea
whose equation of motion is given by
\beq
\left(\Box - \p_5^2 + \frac{\g^2 \vh^2}{4} \right) a = 0
\eeq

In unitary gauge, $\xi_A\rightarrow \infty$, the Goldstone mode decouples and the resulting equations of motion of the remaining fields are
\bea
&\left[\left(\Box-\p_5^2+\frac{\g^2\vh^2}{4}\right)\eta^{\mu\nu}-\p^\mu
\p^\nu \right]A_\mu = 0 &\label{BEOMab}\\
&\left[\Box-\p_5^2 + 2 \muh^2\right]h = 0 & \label{hEOMab}\\
&G = 0 &\label{GEOMab}
\eea

The boundary conditions for the fields $h$, $A_\mu$, and $a$ are determined by
demanding the variation of the boundary action to vanish. Before doing so, the
remaining gauge freedom can be used to eliminate the boundary mixing term
between $A_\mu$ and $A_5$ by adding the boundary gauge fixing
term~\cite{Cacciapaglia:2005pa}
\beq
\label{GFboundary}
S_{GF,b}=\int d^4x (- \frac{1}{2\xi_{A,b}\g^2}\left( \p_\mu A^\mu+\xi_{A,b}A_5\right)^2 ).
\eeq
The variation of Eqs.~(\ref{bulkactionfull}),~(\ref{bulkgaugefix})and~(\ref{GFboundary}) in unitary gauge on the boundary yields
\bea
\delta S_{tot,b}&=&\int d^4 x \left( -\p_5h\delta h + \frac{1}{\g^2}\left[\frac{1}{\xi_{A,b}}\p^\nu\p_\mu A^\mu-\p_5 A^\nu\right]\delta A_\nu \right. \nonumber\\
		&& \left. +\frac{1}{\g^2}\left[- \xi_{A,b}A_5\right]\delta A_5 + a \delta \chi \right).
\label{UEDboundaryvar2}
\eea
Taking $\xi_{A,b}\rightarrow \infty$, and using the definitions of $G$ and $a$
in Eqs. (\ref{Goldstone:eq},\ref{physHiggs:eq}), the boundary conditions
following from Eq.(\ref{UEDboundaryvar2}) correctly reproduce the standard UED
boundary conditions
\bea
\p_5 A_\mu&=&0\nonumber\\
A_5&=&0\nonumber\\
\p_5 h&=&0\nonumber\\
\p_5 \chi&=&0,\label{standardbc}
\eea
which have been used in Ref. \cite{Muck:2001yv} as a starting point to derive the KK-decomposition of the UED model.

In the presence of additional boundary terms the bulk equations of motion Eqs.~(\ref{BEOMab})-(\ref{GEOMab}) are unchanged, but the boundary variations and therefore the boundary conditions on the KK wavefunctions are altered and affect the wavefunctions and mass spectra. We discuss the effects of boundary localized kinetic terms for gauge and the Higgs field in Appendix~\ref{appgaugefixbd}.

\subsection{Generalization to $SU(2)\times U(1)$}\label{NAUED}

It is straightforward to generalize the results of the previous subsection to electroweak sector of UED.  The only complicating factor is that the bulk action Eq.~(\ref{UEDbulkaction}) contains mixing terms of the $U(1)$ and
the electrically neutral $SU(2)$ gauge field $\propto \vh^2 B^\mu W^3_\mu$
due to the Higgs transforming under both $U(1)_Y$ and $SU(2)_L$.

In the absence of any BLTs as discussed here, or if the boundary terms introduced induce the same mixing as the bulk terms, the mixing term can be removed by
rotating into the 5 dimensional $Z_{\mu}, A_{\mu}$ basis given by\footnote{Note that in this basis, the 5 dimensional kinetic terms of $Z_M$ and $A_M$ are canonically normalized, \ie $S\supset -\frac{1}{4}A_{MN}A^{MN}-\frac{1}{4}Z_{MN}Z^{MN}$.}
\bea
Z_M&=&\frac{1}{\sqrt{\gy^2+\gw^2}}\left(W^3_M-B_M\right)\nonumber\\
A_M&=&\frac{1}{\sqrt{\gy^2+\gw^2}}\left(\frac{\gy}{\gw}W^3_M+\frac{\gw}{\gy}B_M \right).\nonumber
\eea

The bulk equations of motion in the $Z_{\mu}, A_{\mu}$ basis are
\bea
&\left[\left(\Box-\p_5^2+\frac{\gw^2\vh^2}{4}\right)\eta^{\mu\nu}-\p^\mu \p^\nu
\right]W^\pm_\mu = 0 & \label{WEOMapA}\\
&\left[\left(\Box-\p_5^2+\frac{(\gy^2+\gw^2)\vh^2}{4}\right)\eta^{\mu\nu}-\p^\mu \p^\nu \right]Z_\mu = 0 &\label{ZEOMapA}\\
& \left[\left(\Box-\p_5^2\right)\eta^{\mu\nu}-\p^\mu \p^\nu \right]A_\mu = 0 &
\label{BEOMapA}\\
& \left[\Box-\p_5^2 + 2 \muh^2\right]h = 0 &\label{hEOMapA}\\
& \left(\Box - \p_5^2 + \frac{\gw^2 \vh^2}{4} \right) a^{\pm} = 0 &
\label{apmEOMapA}\\
& \left(\Box - \p_5^2 + \frac{(\gy^2+\gw^2) \vh^2}{4} \right) a = 0 &
\label{aEOMapA}\\
& G^\pm = G^3=G_Y = 0, \label{GEOMapA}
\eea
where the fields $a, a^{\pm}, G^3, G^{\pm}, G_Y$ are defined by
\bea
a^\pm&=&  \p_5 \chi^\pm +\frac{\vh}{2} W^\pm_5 \nonumber\\
a&=&\p_5 \chi^3+\frac{\vh}{2}Z_5 \nonumber\\
G_{W^\pm}&=&\p_5W_5^\pm+\frac{\gw^2\vh}{2}\chi^\pm \label{su2u1_ga_def}\\
G_Z&=&\p_5Z_5^3+\frac{(\gy^2+\gw^2)\vh}{2}\chi^3 \nonumber\\
G_A&=&\p_5A_5 \nonumber.
\eea

The boundary conditions are determined from the variation of the boundary action. In the absence of any boundary terms other than terms arising from partial integrations used to derive the bulk equations of motion, they read
\bea
\p_5 W^\pm_\mu = 0  &\;\;\;\;\;& W^\pm_5 = 0\nonumber\\
\p_5 Z_\mu = 0  &\;\;\;\;\;& Z_5 = 0\nonumber\\
\p_5 A_\mu = 0 &\;\;\;\;\;& A_5 = 0\nonumber\\
\p_5 h = 0 &\;\;\;\;\; & \p_5 \chi^3 = 0,
\eea
as expected.

\section{UED with boundary localized terms}\label{appgaugefixbd}

The reformulation of the gauge fixing procedure developed in the last section enables us to incorporate the BLTs of Eq.(~\ref{FullBLT}). The aim of this section is to derive the boundary conditions of all fields in the electroweak sector which we use Sec.~\ref{secUEDtheory} in order to calculate the wavefunctions and mass spectra. For illustration we again discuss the abelian model first and then outline the changes when generalizing to the full electroweak sector.

\subsection{The abelian Higgs model in the presence of boundary terms}

Expanding the abelian analog of the action in Eq.~(\ref{FullBLT}) around $\vh$ to quadratic order yields the boundary terms
\bea
S_{BLKT,H}=&& \int d^5 x \left[\delta(y)+\delta(y-\pi R)\right]\times \left(\frac{r_H}{2} \p_\mu h\p^\mu h - \mu_b^2 h^2 \right.\nonumber\\
&&\left.+\frac{r_H}{2}\p_\mu \chi \p^\mu \chi- \frac{r_A}{4\g^2}A_{\mu\nu}A^{\mu\nu}-\frac{r_H \vh}{2}\p^\mu A_\mu\chi-\frac{r_H}{2} \left(\frac{\vh}{2}\right)^2 A_\mu A^\mu\right).\label{HBLKT}
\eea
The brane localized mixing of the gauge field with $\chi$ can be eliminated by using a brane localized gauge fixing term
\bea
S_{GF,b}&=& -\frac{1}{2\xi_{b}\g^2} \int d^4x \left[ \p_\mu A^\mu+\xi_{b}
\left(A_5 + \frac{r_H \g^2 \vh}{2}\chi \right) \right]^2.
\label{UEDboundaryvar3}
\eea
The boundary variation from Eqs. (\ref{UEDboundaryvar2}, \ref{HBLKT}, \ref{UEDboundaryvar3}) in the gauge $\xi_{b}\rightarrow \infty$ yields the boundary conditions
\bea
(\p_5+r_H\Box+2\mu_b^2)h&=&0\\
\left(\p_5+r_A\left[\Box \eta_{\mu\nu}-\p_\mu\p_\nu\right]+r_H\left(\frac{\g\vh}{2}\right)^2\right) A^\nu &=&0\label{Bbdvarapp}\\
A_5 + \frac{r_H \g^2 \vh}{2}\chi &=&0\label{chibdvar}\\
-r_H\Box \chi -\p_5 \chi +\frac{\vh}{2} A_5 &=& 0. \label{b5bdvar}
\eea
The Higgs and gauge boundary conditions are decoupled, and can be used to derive the KK decomposition (see Sec.~\ref{secUEDtheory}).

The boundary conditions for $\chi$ and $A_5$ mix. In the following we show how to identify the physical pseudo scalar mode $a$ consistent with our bulk identifications Eqs. (\ref{Goldstone:eq},\ref{physHiggs:eq}) and the boundary conditions from Eqs. (\ref{chibdvar},\ref{b5bdvar}).
 We start from the bulk equation of motion for $a$ Eq. (\ref{aEOMapA}).
As $G\equiv 0$ in unitary gauge, we know that
\beq
\p_5 A_5\equiv\frac{\g^2\vh}{2}\chi.
\label{B5tochi}
\eeq
Using this equation, which holds for any $y$, we can manipulate Eq. (\ref{B5EOMab2}) to get the bulk equation of motion for $A_5$\footnote{This solution is consistent with the boundary condition on $\chi$ in Eq (\ref{chibdvar}): Imposing Eq. (\ref{B5bdycond1}) on the boundary,  Eq (\ref{chibdvar}) reads
\beq
(\frac{2}{\g^2 \vh}\Box A_5-\p_5 \chi +\frac{\vh}{2} A_5) |_{\rm 0,\pi R} =0.\nonumber
\eeq
Using Eq (\ref{B5tochi}) this becomes
\beq
\left(\Box -\p_5^2 + \frac{\g^2 \vh^2}{4}\right)A_5 = 0 \nonumber
\eeq
\ie the bulk equation of motion (\ref{B5EOM}) which is satisfied everywhere.}
\beq \left(\Box -\p_5^2 + \frac{\g^2 \vh^2}{4}\right)A_5 = 0.
\label{B5EOM}
\eeq
 The boundary variation Eq. (\ref{b5bdvar}) together with Eq. (\ref{B5tochi}) implies boundary conditions
\bea
(A_5 + \frac{r_H \g^2 \vh}{2} \chi^3) |_{\rm 0,\pi R} = 0&& \label{B5bdycond1}\\
\Rightarrow \left\{\begin{array}{lll}
(A_5 + r_H \p_5 A_5) |_{y=\pi R} &=& 0 \\
(A_5 - r_H \p_5 A_5) |_{y=0} &=& 0.
\end{array}
\right.&&\label{B5bdycond2}
\eea

The problem of finding the KK decomposition of $A_5$ is reduced to solving Eq. (\ref{B5EOM}) with boundary conditions (\ref{B5bdycond2}), which again resemble the structure discussed in Sec.\ref{secmassiveBLKT} and can be solved for in the same way. The solution for $a$ is proportional to the solution determined and can be found from it by using Eqs. (\ref{physHiggs:eq},\ref{B5tochi}).The normalization condition for $a$ follows by  collecting the kinetic terms for $A_5$ and $\chi$ from the 5D action and using Eqn.~(\ref{B5tochi})
\bea
S\subset &&\int d^5 x \frac{1}{2\g^2}\p_\mu A_5\p^\mu A_5+\frac{1}{2}\p_\mu\chi\p^\mu\chi[1+r_H(\delta(y)+\delta(y-\pi R))]\nonumber\\
        = &&\int d^5 x \frac{1}{2\g^2}\p_\mu A_5\p^\mu A_5+\frac{1}{2\g^2}\frac{4}{\g^2\vh^2}\p_\mu\p_5 A_5 \p^\mu\p_5A_5[1+r_H(\delta(y)+\delta(y-\pi R))]
\eea
implying the normalization condition for $a$ via
\beq
\delta_{ij}=\frac{1}{\g_I}\int  dy \left\{f^{a}_i f^{a}_j+\frac{1}{\hat{m}_{a}^2}f'^{a}_i f'^{a}_j\left(1+r_H  \left[\delta(y)+\delta(y-\pi R)\right]\right)\right\},
\label{anormcondt}
\eeq
where we defined $\hat{m}_a\equiv \g\vh/2$.

\subsection{Boundary terms and spontaneously broken $SU(2)\times U(1)$}

The main difference compared to the abelian case again arises due to $B_\mu-W^3_\mu$ mixing. In Sec.~\ref{NAUED}, we were able to diagonalize the bulk action because the kinetic terms had a global $U(1)$ symmetry involving the $W_{\mu}^3$ and $B_{\mu}$ fields which was broken by the Higgs VEV. With the addition of the boundary terms, the global U(1) is broken, unless $r_B = r_W$, and hence the basis in which the kinetic terms are diagonal differs from that in which the bulk mass terms are diagonal.

\bigskip

If $r_B=r_W\equiv r_{EW}$, the generalization of the previous section is straightforward. Working in the $Z_M,A_M$ basis defined in Eq.~(\ref{ZAbulkbasis}), the bulk equations of motion for all fields are given in Eqs.~(\ref{WEOMapA}-\ref{GEOMapA}). The boundary terms in this basis read
\bea
S_{BLKT,B}&= \int d^5 x & \left[\delta(y)+\delta(y-\pi R)\right] \left(- \frac{r_{EW}}{4}A_{\mu\nu}A^{\mu\nu} -
\frac{r_{EW}}{4}Z_{\mu\nu}Z^{\mu\nu}- \frac{r_{EW}}{2\gw^2}W^+_{\mu\nu}W^{-\mu\nu}\right. \nonumber\\
&& r_H \left(\frac{1}{2} \p_\mu h\p^\mu h +\frac{1}{2} \p_\mu \chi^a\p^\mu \chi^a -\frac{\vh}{2}\p^\mu W^\pm_\mu\chi^\mp+\left(\frac{\vh}{2}\right)^2 W^+_\mu W^{-\mu}\right.\nonumber\\
&& \left. -\frac{1}{2}\frac{\sqrt{\gy^2+\gw^2}\vh}{2}\p^\mu Z_\mu\chi^3+\left(\frac{(\gy^2+\gw^2)\vh}{2}\right)^2 Z_\mu Z^\mu \right). \label{NABLTZA}
\eea
The boundary conditions for $W_\mu^\pm$, $Z_\mu$, and $A_\mu$ are all decoupled. For $W_\mu^\pm$ and $Z_\mu$ and the associated physical Higgs modes $a^\pm, a^0$, the discussion and results of the previous section apply with the simple replacements $r_A\rightarrow r_{EW}$ and $\g\rightarrow \gw$  or respectively $\g\rightarrow\sqrt{\gy^2+\gw^2}$. Note, that again we find non-flat zero modes for the gauge bosons unless $r_{EW}=r_H$.

For the photon $A_\mu$, no bulk and boundary mass terms are present. The zero mode of the photon is flat and massless, while the higher KK mode masses are reduced in the presence of the boundary kinetic term. Finally, the Higgs boundary conditions are unaltered as compared to the abelian case, so the expressions of the last section hold unmodified.

\bigskip

The general case $r_W\neq r_B$ is discussed in Sec~\ref{secPhenoBneqW}.

\section{Summary of wavefunctions and mass spectra in the electroweak sector}\label{appSpecs}

For convenience, we summarize the mass spectra and wavefunctions for all fields in the electroweak sector in this appendix. For the neutral gauge sector, we give the expansions in the $A_\mu-Z_\mu$ basis which we use in the special case $r_W=r_B$, as well as in the $B_\mu-W^3_\mu$ basis. In the $B_\mu-W^3_\mu$, the $B_\mu-W^3_\mu$ mixing is treated as a mass insertion, \ie after expanding in the $B_\mu-W^3_\mu$ KK basis given here, the off diagonal contributions from $B_\mu-W^3_\mu$ mixing has do be calculated \emph{after} KK decomposing, and the resulting mass matrix has to be diagonalized as outlined in Sec.~\ref{secPhenoBneqW}.

We give all equations in terms of the parameters $r_I, \g_I, m_I$ which are listed in Table~\ref{tabKKsummary} for the respective electroweak fields.

\begin{table}
\begin{center}
\begin{tabular}{|c|c|c|c|}
\hline
 &  {  $m$} &  {  $m_b$} &  {  $r_{\Phi}$} \\
\hline
 {  $W$} &  {  $\gw^2 \vh^2/4$} &  {  $r_H \gw^2
\vh^2/4$} &  {  $r_W$} \\
\hline
 {  $Z$} &  {  $(\gw^2 + \gy^2) \vh^2/4$} &  {
$r_H (\gw^2 + \gy^2) \vh^2/4$} &  {  $r_B$} \\
\hline
 {  $h$} &  {  $\sqrt{2}\hat{\mu}$} &  {  $\sqrt{2}\mu_b$} &
 {  $r_H$} \\
\hline
 {  $a_{\pm}$} &  {  $\gw^2 \vh^2/4$} &  {  $r_H
\gw^2 \vh^2/4$} &  {  $r_H$} \\
\hline
 {  $a_0$} &  {  $(\gw^2 + \gy^2) \vh^2/4$} &  {
  $r_H (\gw^2 + \gy^2) \vh^2/4$} &  {  $r_H$} \\
\hline
\end{tabular}
\end{center}
\caption{Substitutions needed to convert the results of the 5 dimensional
massive scalar in Section~\ref{secmassiveBLKT} in those for the electroweak
fields.}
\label{tabKKsummary}
\end{table}

The general form of the wavefunctions is given by

\bea
f^I_\alpha=N^I_\alpha \left\{\begin{array}{cc}
\frac{\cosh \left(M_{I^{\alpha}} \left(y-\frac{\pi R}{2}\right)\right)}{\cosh \left(
\frac{M_{I^{\alpha}} \pi R}{2}\right)} & \alpha \mbox{ even} \\
-\frac{\sinh \left(M_{I^{\alpha}} \left(y-\frac{\pi R}{2}\right)\right)}{\sinh \left(
\frac{M_{I^{\alpha}} \pi R}{2}\right)} & \alpha \mbox{ odd} \\
\end{array} \right.
\eea
\bea
f^I_n=N^I_n \left\{\begin{array}{cc}
\frac{\cos \left(M_{I^{(n)}} \left(y-\frac{\pi R}{2}\right)\right)}{\cos \left(
\frac{M_{I^{(n)}} \pi R}{2}\right)} & n \mbox{ even} \\
-\frac{\sin \left(M_{I^{(n)}} \left(y-\frac{\pi R}{2}\right)\right)}{\sin \left(
\frac{M_{I^{(n)}} \pi R}{2}\right)} & n \mbox{ odd} \\
\end{array} \right.,
\eea
where
\bea
m_{I^{(\alpha)}}^2&=&-M_{I^{(\alpha)}}^2+m_I^2\nonumber\\
m_{I^{(n)}}^2&=&M_{I^{(n)}}^2+m_I^2,
\eea
The mass determining equations are
\bea
b^I_\alpha &=& \left\{ \begin{array}{cc}
\tanh (\frac{\sqrt{m_I^2-m_{I^{(\alpha)}}^2} \pi R}{2}) & \alpha \mbox{ even} \\
\coth (\frac{\sqrt{m_I^2-m_{I^{(\alpha)}}^2} \pi R}{2}) & \alpha \mbox{ odd}
\end{array}  \right.  \nonumber\\
b^I_n &=& \left\{ \begin{array}{cc}
-\tan (\frac{\sqrt{m_{I^{(n)}}^2 - m_I^2} \pi R}{2}) & n \mbox{ even} \\
\cot (\frac{\sqrt{m_{I^{(n)}}^2 - m_I^2} \pi R}{2}) & n \mbox{ odd}.
\end{array} \right. \label{nQcondHBLKT}
\eea
where $m_{I^{(i)}}^2$ are the \emph{physical} KK masses, and
\bea
b^h_i=\frac{r_H m^2_{h^{(i)}}-2\mu_b^2}{\sqrt{|m_{h^{(i)}}^2-m_h^2|}}\;\;\;\;
b^G_i=\frac{r_G m^2_{G^{(i)}}-r_H m_G^2}{\sqrt{|m_{G^{(i)}}^2-m_G^2|}}\;\;\;\;
b^{a^{\pm,0}}_i=r_H\sqrt{|m_{a^{\pm,0(i)}}^2-m_{a^{\pm,0}}^2|}.
\label{bHabelian}
\eea
where $i,j\in\{\alpha,n\}$ for the Higgs $h$, the additional Higgs degrees of freedom $a^{\pm,0}$, any gauge fields $G\in(W^{\pm,3},B,Z,A)$.

The normalization conditions are
\bea
\delta_{ij}&=&\int  dy f^h_i f^h_j\left(1+r_H  \left[\delta(y)+\delta(y-\pi R)\right]\right)\\
\delta_{ij}&=&\frac{1}{\g_I^2}\int  dy f^{\tilde{G}}_i f^{\tilde{G}}_j\left(1+r_{\tilde{G}}  \left[\delta(y)+\delta(y-\pi R)\right]\right)\\
\delta_{ij}&=&\int  dy f^{Z,A}_i f^{Z,A}_j\left(1+r_{EW}  \left[\delta(y)+\delta(y-\pi R)\right]\right)\\
\delta_{ij}&=&\frac{1}{\g_I^2}\int  dy \left\{f^{a^{\pm,0}}_i f^{a^{\pm,0}}_j+\frac{1}{\hat{m}_{a^{\pm,0}}^2}f'^{a^{\pm,0}}_i f'^{a^{\pm,0}}_j\left(1+r_H  \left[\delta(y)+\delta(y-\pi R)\right]\right)\right\},
\eea
where $\tilde{G}$ denotes gauge fields $G\in(W^{\pm,3},B)$. The difference in normalizations in the $W-B$ basis and the $Z-A$ basis of  Eq.~(\ref{ZAbulkbasis}) arises because in the latter, the 5 dimensional kinetic terms are canonically normalized already, \ie $S\supset -\frac{1}{4}A_{MN}A^{MN}-\frac{1}{4}Z_{MN}Z^{MN}$. The normalization conditions for $a^{\pm,0}$ follow as in Eq.~(\ref{anormcondt}).

\end{document}